\documentclass[amssym]{mn2e}

\usepackage{graphicx}
\usepackage{subfigure}
\def\pageoffset#1#2{\hoffset=#1\relax\voffset=#2\relax} 
\pageoffset{0pc}{2.5pc} 
 at 10truept

\def\and  {\it {et al.} \rm}

\def\spose#1{\hbox to 0pt{#1\hss}}
\def\simlt{\mathrel{\spose{\lower 3pt\hbox{$\mathchar"218$}}
     \raise 2.0pt\hbox{$\mathchar"13C$}}}
\def\simgt{\mathrel{\spose{\lower 3pt\hbox{$\mathchar"218$}}
     \raise 2.0pt\hbox{$\mathchar"13E$}}}
\def\be{\begin{equation}}
\def\ee{\end{equation}}
\def\bce{\begin{center}}
\def\ece{\end{center}}
\def\bea{\begin{eqnarray}}
\def\eea{\end{eqnarray}}
\def\ben{\begin{enumerate}}
\def\een{\end{enumerate}}

\def\ni{\noindent}

\def\brr{\begin{array}}
\def\err{\end{array}}

\def\btau{\overline{\tau}}

\def\nh1{n_{\rm HI}}

\def \p1dk {P_{\rm 1D}(k)}
\def \simlt {\mathrel{\spose{\lower 3pt\hbox{$\mathchar"218$}}
     \raise 2.0pt\hbox{$\mathchar"13C$}}}
\def \simgt {\mathrel{\spose{\lower 3pt\hbox{$\mathchar"218$}}
     \raise 2.0pt\hbox{$\mathchar"13E$}}}

\def \P {{\cal P}}

\def \vr {{\bf r}}

\def \vH {{\cal H}}

\def \be {\begin{equation}}
\def \en {\end{equation}}
\def \bea {\begin{eqnarray}}
\def \ena {\end{eqnarray}}

\def \bi {\begin{itemize}}
\def \ei {\end{itemize}}

\def \apj {{\it ApJ}}
\def \apjl {{\it ApJ Let.}}

\def \mnras {{\it MNRAS}}

\usepackage[usenames]{color}
\definecolor{Blue}{rgb}{0,0.08,0.65}
\definecolor{Red}{rgb}{0.65,0.08,0.05}
\definecolor{Green}{rgb}{0.15,0.45,0.25}
\def\blue{\color{Blue}}

\def\Xtophe#1{\noindent{ \blue\bf[$\spadesuit$ #1]}}

\begin{document}
\title[The three dimensional skeleton: tracing the filamentary structure of the universe]{
The three dimensional skeleton:\\ tracing the filamentary structure of the Universe.}

\author[T. Sousbie, C. Pichon, S. Colombi, D. Novikov  \& D. Pogosyan ]
{\ni T. Sousbie$^{1,2}$, C. Pichon$^{1,2}$, S.  Colombi$^1$, D. Novikov$^3$ \&  D. Pogosyan$^4$  \\
$^1$ Institut d'Astrophysique de Paris \& UPMC,  98 bis boulevard Arago, 75014 Paris, France \\
$^2$ Centre de Recherche Astrophysique de Lyon,  9 avenue Charles Andr\' e, 69561 Saint Genis Laval, France \\
$^3$ Astrophysics, Blackett Laboratory, Imperial College London, London SW7 2AZ, England\\
$^4$ Department of physics, University of Alberta, 412 Avadh Bhatia Physics Laboratory, Edmonton, Alberta, T6G 2J1, Canada  \\
sousbie@iap.fr, pichon@iap.fr, colombi@iap.fr, pogosyan@phys.ualberta.ca, novikov@astro.ox.ac.uk
}

\maketitle

\begin{abstract}
  
The skeleton formalism aims at extracting and quantifying the filamentary structure of the universe is generalized to 3D density fields;  
 a numerical method  for
computating  a local approximation of the skeleton is presented and validated here 
on  Gaussian random fields.
It involves solving equation $\, (\,\vH\,\nabla \rho \times \nabla \rho\,)=0$ where $\nabla \rho$ and $\vH$ are the gradient and Hessian matrix of the field.
This method manages to trace well the filamentary structure in 3D fields such as given by 
numerical simulations of the  dark matter distribution on large
scales and is insensitive to monotonic biasing.  \\
Two of its characteristics, namely  its length
and differential length, are analyzed for Gaussian random
fields.  Its differential length per unit normalized density 
contrast  scales like the PDF of the underlying density contrast  times the total length
times a quadratic Edgeworth correction involving the square of the spectral parameter.
The total length scales like the inverse square smoothing length,
 with a scaling factor given by $0.21 (5.28+ n)$ where $n$ is the 
power index of the underlying field.
This dependency implies that the total length can be used to constrain 
the shape of the underlying power spectrum, hence the cosmology.\\
Possible applications of the skeleton to galaxy formation and cosmology are discussed.
As an illustration, the orientation of  the spin of dark halos and the orientation 
of the flow near the skeleton is computed for dark matter simulations. 
The flow is laminar along the filaments, while  spins of dark halos within $500$ kpc of the skeleton are preferentially orthogonal
to the direction of the flow at a level of $25\%$.
\end{abstract}
\section{Introduction}

Recent galaxy surveys like 2dF (Colless \& al.2003) or SDSS (Gott \& al. 2005) emphasized the complexity of the
matter  distribution in the universe which presents  large scale structures such as
filaments,  clusters or walls on the boundaries of low density bubbles
(voids).   
On the theoretical side, the currently  favoured
scenario suggests that the universe evolved from
Gaussian initial  conditions to form the
structures that are observed nowadays. Numerical simulations
have successfully --both statistically and visually,  captured the main features of  the observed filamentary distribution.

Novikov,
Colombi \& Dore 2005 (NCD)  introduced  the skeleton formalism in 2D, which aims at extracting and analysing the
filamentary structure of a given density field.
This paper extends it to three dimensions in order to describe the universe's  large scale matter
distribution and its dynamical environment.

In the literature, various steps towards a quantitative description of the large  structures have  been suggested.
Statistical tools such as  correlation functions (e.g., Peebles 1980) and 
power spectra (e.g. Peacock  1998) have been widely used and
have been  successful in describing matter distribution and constraining cosmological parameter.
Recently, fast algorithms have been designed for first and second order (Szapudi \& al. 2005), as well as higher order  statistics (counts in cells etc..) as in (Croton \& al. 2004) or (Kulkarni \& al. 2007). 
Minkowski  functionals (see, e.g., Kerscher 2000 for a review) are so-called ``shape
finders'' (Sahni \& al. 1998) attempt instead to describe the  topology of a distribution (see  (Sheth \& Sahni 2005) for application to large scale structure)
and can discriminate between Gaussian and non  Gaussian fields  as shown in (Doroshkevich \& al. 1970), (Gott \& al. 1986)  or more recently (Hikage \& al. 2006)). 
These topological and statistical estimators analyse the distribution of observed galaxies globally and uniformly, and make little attempt at recovering the precise geometry of the matter distribution, i.e. they do not focus on specific regions (such as clumps, voids and filaments).

Focusing on the identifiable regions of  the universe, 
 the peak patches theory (Bond \& Myers 1996) 
attempts to describe cosmic structures formation through the
identification of  the collapse of the dense regions near the density peak and surrounding patches. In this
framework, the evolution of patches hierarchy can be understood from
the measurement of only a few characteristics of the patches, while assuming that their
flow does not depend on their internal non-linear dynamics.
This line of thought has been extended in the Cosmic Web paradigm (Bond, Kofman, Pogosyan, 1996), which has emphasized
that the large scale spatial distribution of galaxy clusters and the filaments between them can be understood as mildly non-linear 
enhancement  of high density peaks and filamentary ridges present in the initial gaussian density field. 
 Recently, Hanami (2001) presented  the so-called skeleton tree formalism:
it  analyses the process of hierarchical merging and extends
the language  of the peak patch  through the analysis of the 
ridges of the density field in an abstract space corresponding to 
the usual three dimensions augmented by the smoothing length. 

The structure of voids in  the large scale dark matter distribution also has an extended history
of theoretical modeling  (see e.g. Hoffman
\& Shaham 1982, Icke 1984 or Bertschinger 1985) while various void
identifiers have been designed (see e.g. Platen 2007 and references
therein).

One of the first attempts to develop an algorithm to detect and trace the filaments in the particle distribution 
has been the minimal spanning tree (MST)  technique proposed by Doroshkevich in (Doroshkevich \& al 1970,
Barrow \& al 1985). Starting from a  point distribution (a galaxy
survey or a dark matter simulation),  this  method constructs the graph that connects all the dots with  the property of
never forming closed paths and being of minimal total lengths.
Interesting statistical features can be extracted from it like
the shape of the clusters or the length of the trunk (the longest
path) and branches which are characteristic of the filamentarity of
the distribution.\\

The three-dimensional skeleton described  in this paper focuses on the 
critical lines of a distribution, i.e. the set of lines joining the critical points 
 in order to be  able to
compute the characteristic features of the underlying field (such as
the total length of the filaments in a cosmological dark matter
distribution).  The skeleton provides a simple mathematical definition of
the filaments of a density field based on Morse theory (see, e.g.,
Milnor 1963;  Colombi, Pogosyan \& Souradeep 2000; Jost 2002, Novikov et al. 2006) and thus
allows their extraction as well as their characterisation.


Section 2 defines the local skeleton of large scale structures.
Section~3 introduces the  numerical algorithm for constructing the local
skeleton, and  discusses its properties  near the critical points 
(Appendix A gives a more detailed description of the algorithm). 
Section~4  investigates the 
evolution of its  differential and total length as a 
function of the properties of the underlying field.
Appendix C sketches the derivation of this differential length.
Possible applications to cosmology and galaxy formation
 are 
discussed in section~5, where two illustrations regarding the 
nature of the dark matter flow near the skeleton are given.

\section{The local skeleton: theory}
\label{sec:3dskl}
A comprehensive definition of  the skeleton and how its local
approximation in two dimensions is derived can be found in Novikov,
Colombi \& Dore 2006. To sum up, the so-called ``real'' skeleton is by
definition the subset of critical lines joining the saddle points of a field to
its maxima while following the gradient's direction (while critical lines 
link {\sl all} kinds of critical points together). It is easy to picture that
applying this definition to a 2D field (an altitude map in a mountainous
region for instance) allows the extraction of the ridges of that distribution. Although
simple in appearance, this definition presents the drawback that it is
in essence non-local: the presence of the skeleton in a given
sub-region may depend on the presence of a saddle point in a different
sub-region.
In order to enforce locality,
an approximation can in fact be derived using Taylor expansion in
the vicinity of the critical points (i.e. local maxima and saddle
points), leading to a second order approximation of the skeleton: the
{\rm local} skeleton.

\subsection{The 2D local skeleton}

 Defining the  {\it local critical lines }
 as the set of points where the gradient
of the field is an extremum along an isodensity contour, it can be
shown (Novikov,
Colombi \& Dore 2006) that this set of points obeys the equation:
\begin{eqnarray}
  {\cal S} & \equiv & \frac{\partial \rho}{\partial r_1} \frac{\partial
    \rho}{\partial r_2} \left( \frac{\partial^2 \rho}{\partial r_1^2} -
  \frac{\partial^2 \rho}{\partial r_2^2} \right)  \nonumber \\& +  &
   \frac{\partial^2
    \rho}{\partial r_1 \partial r_2}\left( \left[ \frac{\partial \rho}{\partial r_2} \right]^2 - \left[\frac{\partial
      \rho}{\partial r_1} \right]^2
  \right)=0,
  \label{eq:2dsdef}
\end{eqnarray}
where $r_1$ and $r_2$ denote space coordinates and $\rho(r_1,r_2)$ is the
density field. Equation~(\ref{eq:2dsdef}) can be rewritten 
\begin{equation}
{\cal S} = {\rm det}\, (\,\vH\,\nabla \rho, \nabla \rho\,)=0, \label{eq:2dlocalsketot}
\end{equation}
where $\vH\equiv{\partial^2 \rho}/{\partial r_1\partial r_2}$ is the
Hessian (second derivatives matrix) of the field. This can be
interpreted mathematically as the set of points where the
gradient of the field is an eigenvector of the Hessian (that is,
gradient and main curvature axis are aligned), which is clearly a
local property of the field.\\

However, in order to correspond to the ``real'' skeleton that 
 traces the ridges  and clumps of the field (its structure), another
condition has to be enforced.  For  its local approximation, it is
equivalent to stating that the gradient should be {\sl minimal} (every point
of the local skeleton of coordinates $\vr$ should also be a local {\rm
minimum} of the isodensity contour at density $\rho\left(\vr
\right)$). That is, one has to enforce the condition that
the second eigenvalue of the Hessian should be negative:
\begin{equation}
  \begin{array}{l}
  \displaystyle \lambda_2 < 0 \,, \quad {\rm and } \quad
  \displaystyle \vH \,\nabla\rho=\lambda_1 \nabla\rho\,,\\
  \end{array}
\label{eq:eigselect2D}
\ee  
where $\lambda_i$ are the eigenvalues of the Hessian and $\lambda_{2}<\lambda_{1}$.

\subsection{The 3D local skeleton}
\begin{figure}
  \centering
  \includegraphics[angle=0,width=8cm]{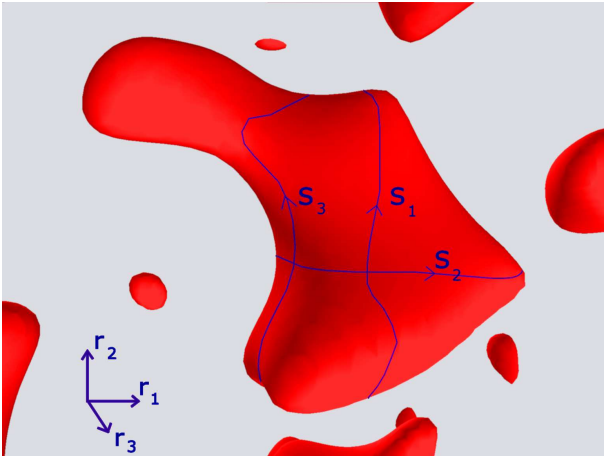}
  \caption{Definition of the coordinate system on an isocontour.}
\label{fig:conecut}
\end{figure}

Let us now derive the generalization of the notion of  the local skeleton to a
three-dimensional space. The philosophy is essentially the same but minor
differences arise which are addressed here. 

Starting from the same definition as in 2D, the 
skeleton should be the set of points where the density is an extremum
along an isodensity contour. Let $(u,v)$ be a coordinate system along
an isocontour
$\left(r_1\left(u,v\right),r_2\left(u,v\right),r_3\left(u,v\right)\right)$
where $r_i,i\in\lbrace1..3\rbrace$ are the three space coordinates. The
 definition of an isocontour implies that :
\begin{equation}
  \left\lbrace
  \displaystyle\begin{array}{l}
  \displaystyle\frac{\partial \rho}{\partial r_1} \frac{d r_1}{d u} + 
  \frac{\partial \rho}{\partial r_2} \frac{d r_2}{d u} + 
  \frac{\partial \rho}{\partial r_3} \frac{d r_3}{d u} = 0\ , \\ \
  \displaystyle\\
  \displaystyle\frac{\partial \rho}{\partial r_1} \frac{d r_1}{d v} +
  \frac{\partial \rho}{\partial r_2} \frac{d r_2}{d v} + 
  \frac{\partial \rho}{\partial r_3} \frac{d r_3}{d v} = 0.\,
  \end{array}
  \right .
  \label{eq:3disocontour}
\ee

Moreover, as the gradient of the field $\rho$ has to be an extremum:
\begin{equation}
  \begin{array}{l}
     \displaystyle\frac{d}{du}(|\nabla \rho|^2)=0\,, \quad {\rm and}\quad
     \displaystyle\frac{d}{dv}(|\nabla \rho|^2)=0
  \end{array}.
  \label{eq:3dgradnull}
\end{equation}
Using equations~(\ref{eq:3disocontour}) and (\ref{eq:3dgradnull}), let
us derive the equation of the local critical lines, which should
only depend on the field and its first and second order spatial
derivatives, similarly to equation~(\ref{eq:2dsdef}). To do so, a coordinate system along the
isocontour is needed but, as opposed to the 2D case, any
coordinate system defined on an isocontour will be singular in some
place as the isocontour is a closed surface. In order to avoid this problem, we choose to
define three coordinates systems and swap from one to another when it
becomes singular.\\

Defining $s_i$ three one-dimensional coordinates
systems so that for different values of $s_i$, one remains in the plane
$(\bf{r_j},\bf{r_k})$ where $i \neq j \neq k$ and $i,j,k \in \lbrace1..3\rbrace$. The
coordinates system $s_i$ is singular wherever $\nabla \rho\,
\alpha\,\bf{r_i}$. The constrain is  to satisfy equations.~(\ref{eq:3disocontour}) and
(\ref{eq:3dgradnull}) for $u\equiv s_i $ and $v\equiv s_j $ with $i\neq j$. For
any $s_i$, these read:
\begin{equation}
  \begin{array}{l}
     \displaystyle\frac{d}{d{s_i}}(|\nabla \rho|^2)=0 \,,\,\,\,{\rm and}\,\,\,\,
     \displaystyle\frac{\partial \rho}{\partial r_1} \frac{d r_1}{d s_i} + 
     \frac{\partial \rho}{\partial r_2} \frac{d r_2}{d s_i} + 
     \frac{\partial \rho}{\partial r_3} \frac{d r_3}{d s_i} = 0
  \end{array}.
  \label{eq:3dgradnullrewrote}
\end{equation}
Choosing $i \neq j \neq k \in \lbrace1..3\rbrace$, this system becomes after some algebra:
\begin{equation}
 \displaystyle
\begin{array}{lll}
 \displaystyle{\cal{S}}_i & \displaystyle\equiv &\displaystyle \frac{\partial^2 \rho}{\partial r_j\partial r_k} 
\left( \frac{\partial \rho}{\partial r_j}^2 - \frac{\partial \rho}{\partial
  r_k}^2 \right)\\
 \displaystyle&\displaystyle+& \displaystyle\frac{\partial \rho}{\partial r_j} \frac{\partial \rho}{\partial r_k}
\left( \frac{\partial^2 \rho}{\partial r_k^2} - \frac{\partial^2
  \rho}{\partial r_j^2}\right)\\
 \displaystyle&\displaystyle-&\displaystyle \frac{\partial \rho}{\partial r_i} \left( \frac{\partial \rho}{\partial
  r_k} \frac{\partial^2 \rho}{\partial r_i\partial r_j} - \frac{\partial
  \rho}{\partial r_j} \frac{\partial^2 \rho}{\partial r_i\partial r_k} \right)
= 0\,.
\end{array}
\label{eq:3dSi}
\end{equation}
Indeed, equation (\ref{eq:3dSi}) reduces equation (\ref{eq:2dsdef}) in the 2D case,
when assuming that the field is constant in the direction orthogonal to that 2D
plane (the first two terms of equation (\ref{eq:3dSi}) are the same as in
equation~(\ref{eq:2dsdef})).
The local critical lines are thus the set of points that satisfies:
\begin{equation}
{\bf{{\cal{S}}}} \equiv
\left( 
\begin{array}{c}
{\cal{S}}_i\\
{\cal{S}}_j
\end{array}
\right)
\,=\,{\bf{0}} ,\; i\neq j\in{\lbrace1,2,3\rbrace}.
\label{eq:3dS_0}
\end{equation}
It is interesting to note that, as in the 2D case, equation
(\ref{eq:3dS_0}) defines the local critical line as the set of points where the
gradient of the density is an eigenvector of its Hessian matrix (the
gradient and the principal curvature axis are collinear):
\begin{equation}
{\cal S}=\, (\,\vH\cdot \nabla
\rho \times \nabla \rho\,)=\mathbf{0}. \label{eq:defS2}
\end{equation}
Once again,  in order to require that the skeleton traces only the ridges of the
distribution (i.e. the filaments in 3D),
retrieving the subset of local critical lines that
define the local skeleton can be achieved by enforcing  a negativity
condition on the weakest eigenvalues of the Hessian:
\begin{equation}
  \begin{array}{l}
  \displaystyle \lambda_2 < 0, \quad
  \displaystyle \lambda_3 < 0, \quad
  \displaystyle \vH \,\nabla\rho=\lambda_1 \nabla\rho \label{eq:eigen} \\
  \end{array}.
\label{eq:eigselect3D}
\end{equation}  
That is, the local skeleton is the subset of the local critical where
the norm of the 3D gradient is {\rm minimal} along the 2D isodensity
contours (as opposed to  simply extremal).
Note that from equation~(\ref{eq:3dS_0}) it is straightforward to show that any monotonic function 
of the field will have exactly the same skeleton as the field itself.

\section{Implementation and features}

\subsection{Implementation}

Equation~(\ref{eq:3dS_0}) is at the basis of the numerical
implementation of the local skeleton determination
developed here. The details of the algorithm are described in   Appendix A, while the optimal choice of  resolution and
smoothing is presented in Appendix B.
All the computations were performed using a specially developed
\texttt{C} package: {\bf \texttt{SkelEx}}\footnote{Available on
request from the authors.}  ({\bf Skel}eton {\bf Ex}tractor). This
package also includes a flexible OpenGL visualization tool that was
used for making the figures in this paper.\\  Figure \ref{fig:skl3D}
presents  the skeleton obtained for a density field sampled from a
numerical simulation of dark matter distribution on a $50 h^{-1}$Mpc
box with $512^3$ particles using GADGET-2 (Springel 2005). The lighter
colors represent denser regions and the blue skeleton appears to match
quite well what one could identify as  the filaments by eye. Note that
the skeleton is both a tracer of the topology (it links a sub-set of
the critical points) and the geometry of its underlying density. Hence
it can be used to compare the geometrical  and topological properties
of various fields, e.g. the temperature and  the dark matter
distribution in hydro-dynamical simulations.
See also Figure~A3 for a graphical description of how the local skeleton is drawn.

\begin{figure*}
\centering
\includegraphics[width=15cm,height=12cm]{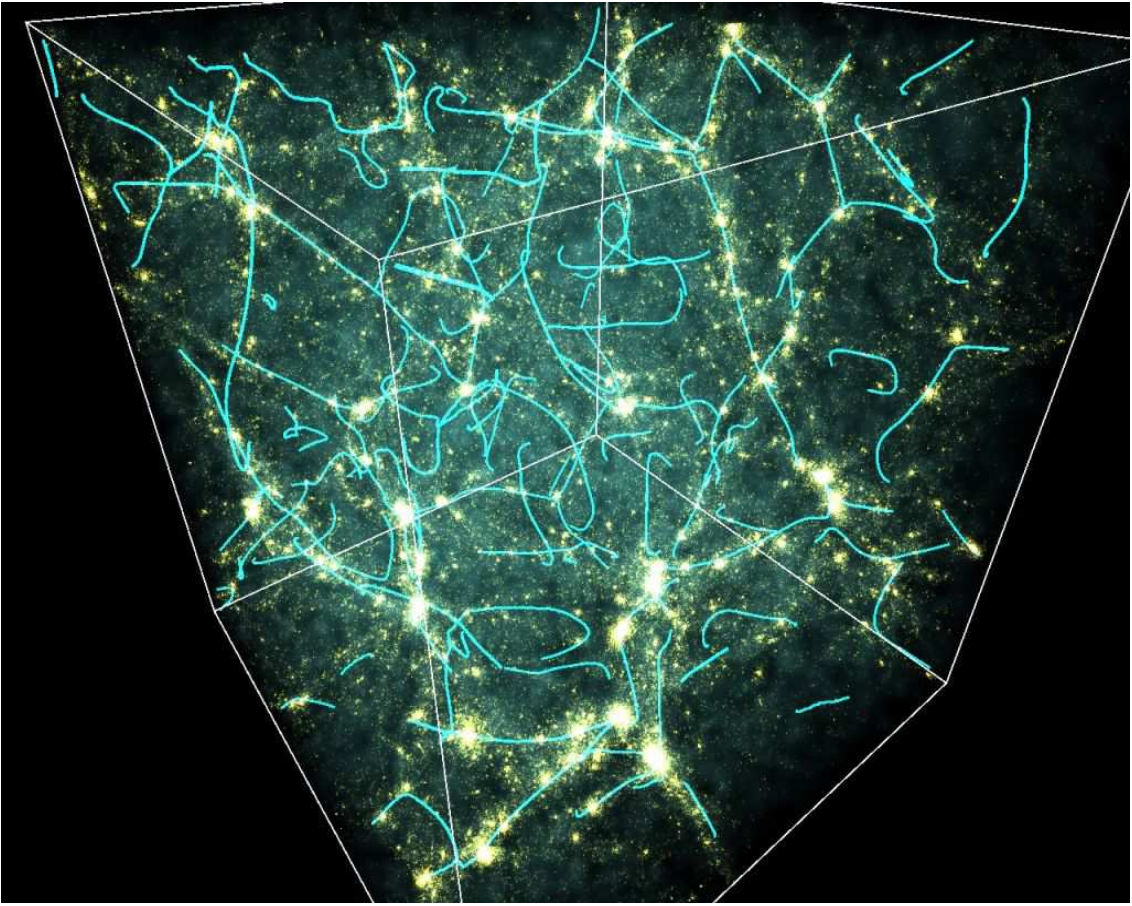}
\caption{ The final 3D skeleton derived from a $50$ Mpc standard
$\Lambda$CDM simulation run with Gadget-2 using $512^3$
particles. This result is obtained after post treating the skeleton
using the method described in Appendix \ref{imple}. \label{fig:skl3D}}
\end{figure*}

\subsection{The local skeleton branching properties}

\begin{figure}
\centering \subfigure[$0>\lambda_1\ge\lambda_2\ge\lambda_3$ $(I=0)$]{\includegraphics[width=4cm]{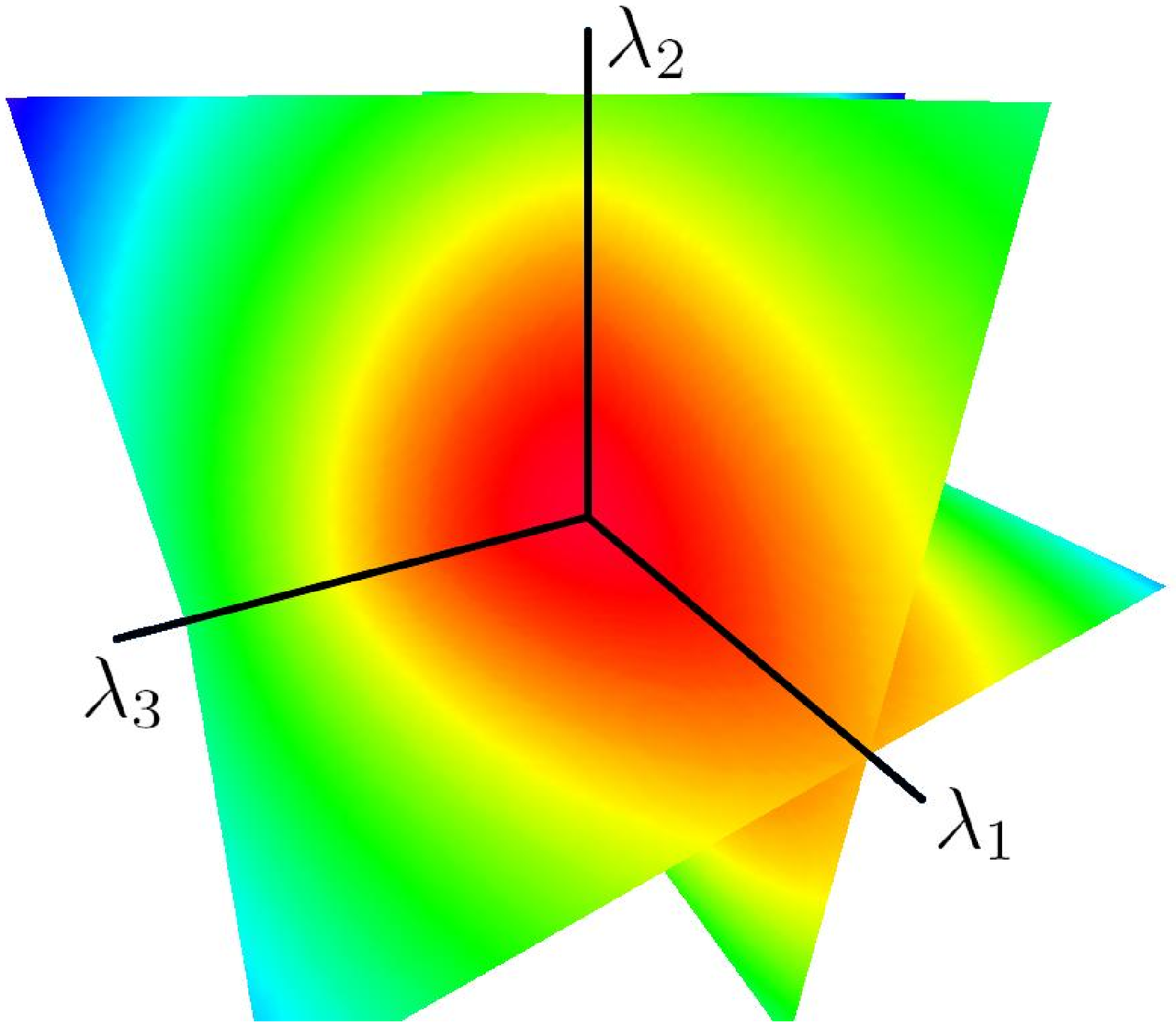}\label{fig_courbure3D_mmm}}
\hfill \subfigure[$\lambda_1>0>\lambda_2\ge\lambda_3$ $(I=1)$]{\includegraphics[width=4cm]{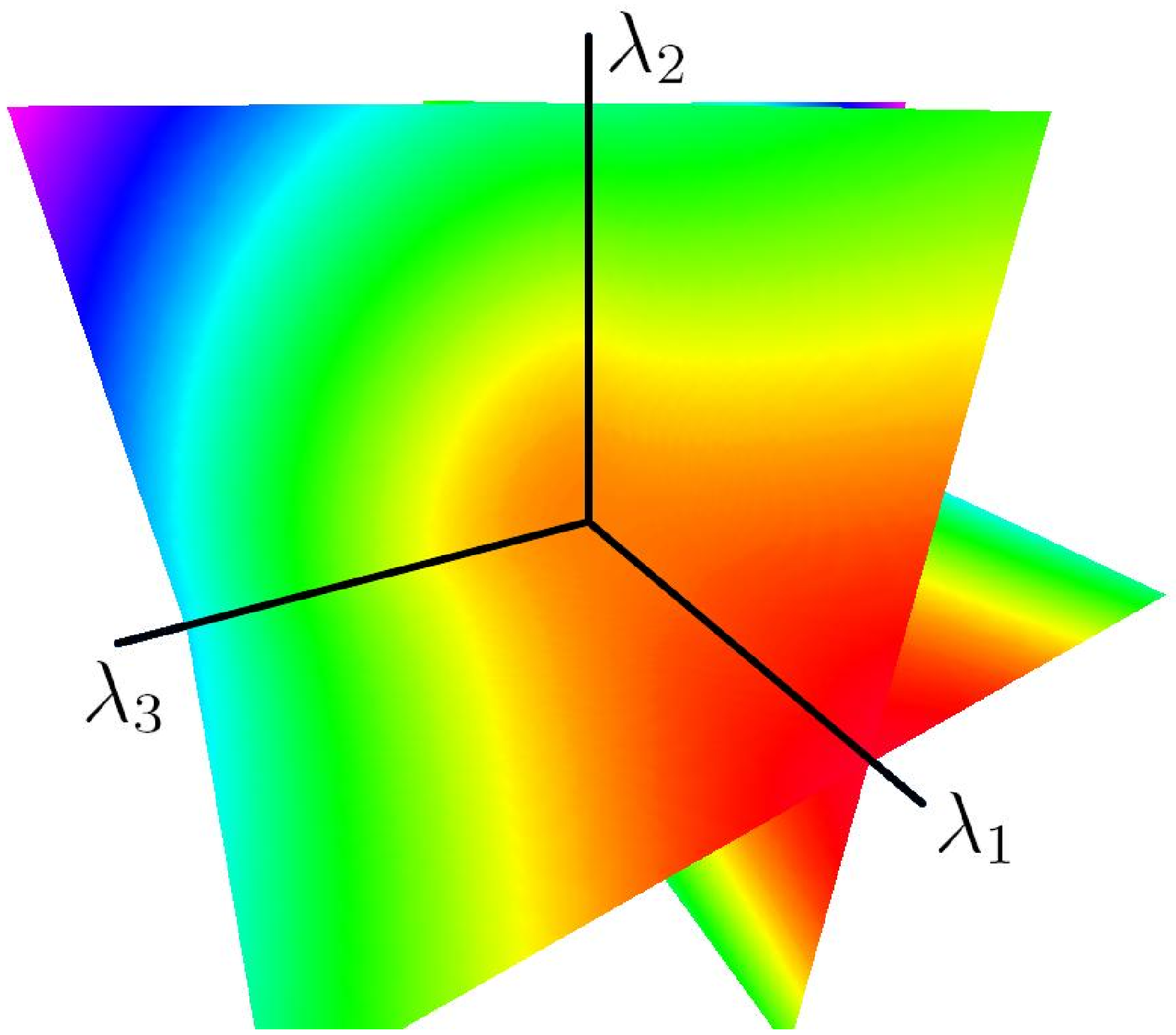}\label{fig_courbure3D_pmm}}\\
\subfigure[$\lambda_1\ge\lambda_2>0>\lambda_3$ $(I=2)$]{\includegraphics[width=4cm]{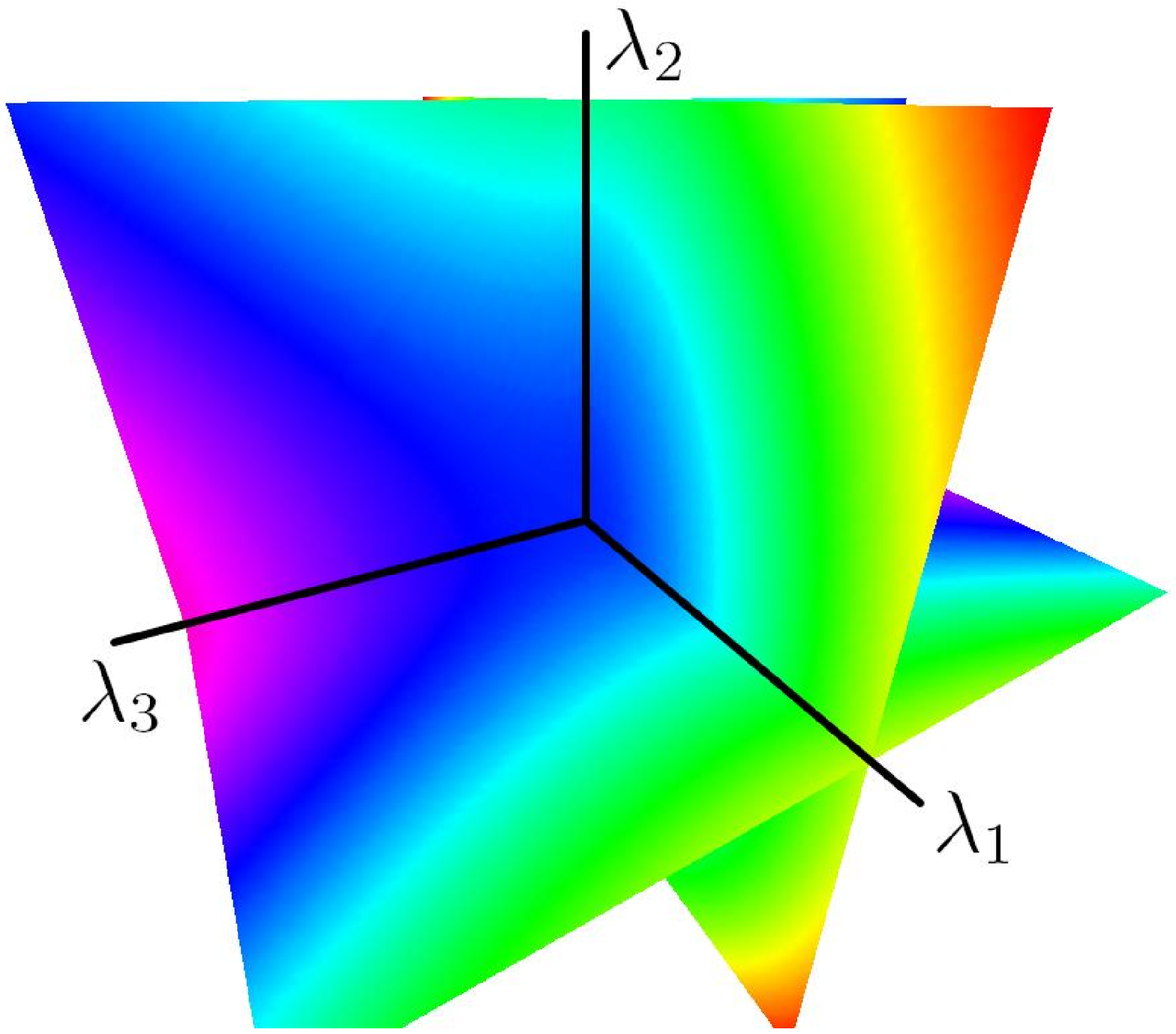}\label{fig_courbure3D_ppm}}
\hfill \subfigure[$\lambda_1\ge\lambda_2\ge\lambda_3>0$ $(I=3)$]{\includegraphics[width=4cm]{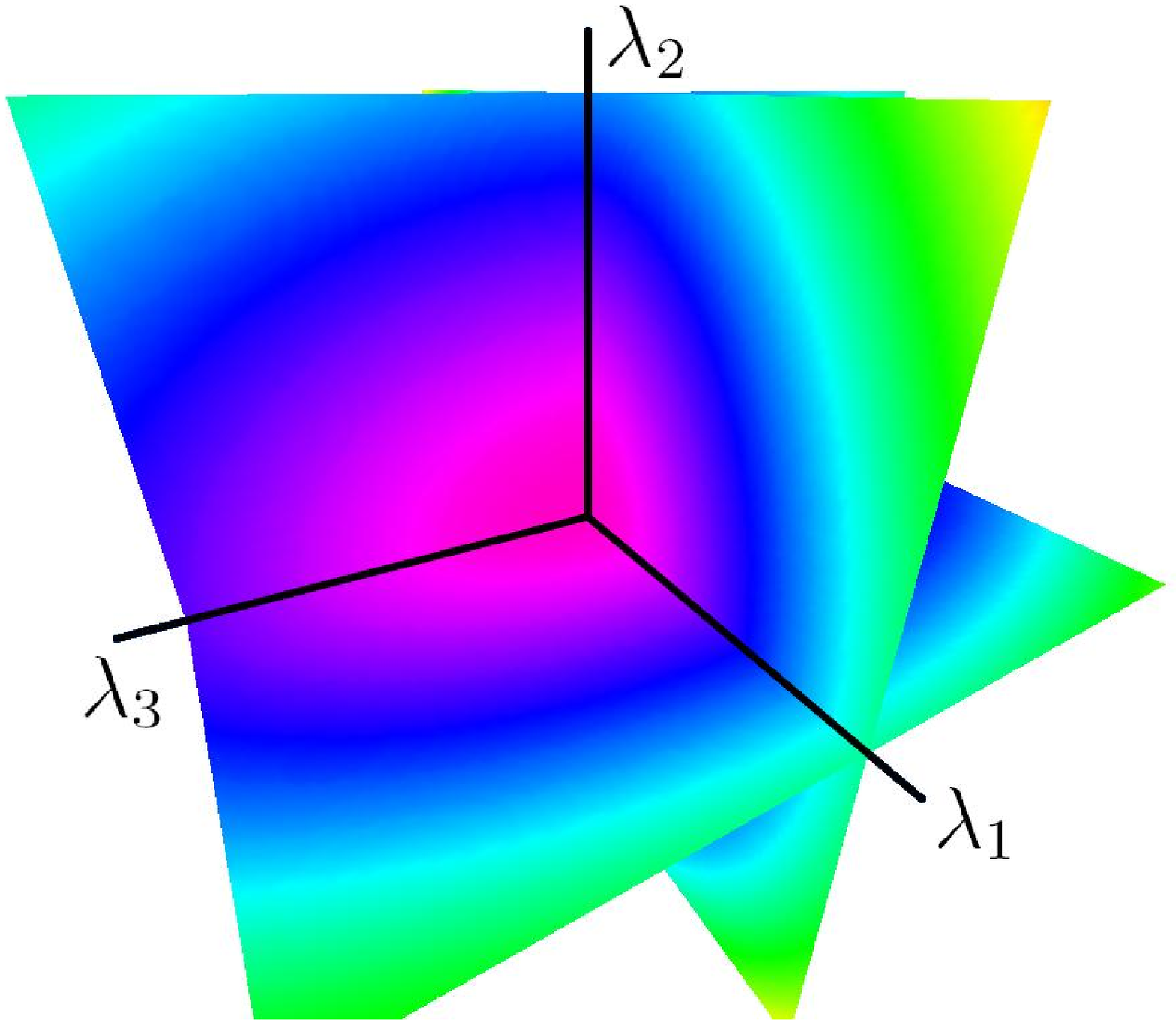}\label{fig_courbure3D_ppp}}
\caption{Illustration of a second-order approximation of the density
field around a maximum ($I=0$), filament ($I=1$) and pancake ($I=2$)
saddle point and a minimum ($I=3$). The color stands for the density,
ranging from purple in low density regions to red in high density
regions. The axes are the eigenvectors of the Hessian, and give
the direction of the 6 branches of the local critical lines  going through
these critical points (i.e. where the gradient of the field and the eigenvectors of $\vH$
are aligned). The skeleton is the subset  of these critical lines linking
maxima (fig. \ref{fig_courbure3D_mmm}) and filament saddle points
(fig. \ref{fig_courbure3D_pmm}), in the direction of the eigenvector
associated with $\lambda_1$.}{\label{fig_courbure3D}}
\end{figure}

Let us now describe some global branching properties of the critical lines and the local skeleton.
Important ingredients of the skeleton are  the extrema of the field.
Indeed, the ``real skeleton'' is defined as a set of critical lines
that connect maxima to saddle points.  Much of the topological
behaviour of the skeleton is related to the distribution of such
extremal points. For the {\em local} skeleton  described  this paper, the role of 
the extrema is similar but the whole set of critical lines encompass aditionnal branches linking all kind
of field extrema together.\\

Since the local skeleton is based on a local second order
approximation of the density field, $\rho$,  its properties can be
understood through the properties of the gradient $\nabla \rho$ and
Hessian matrix $\vH\left( \rho\right)$ only. The eigenvalues of $\vH$
define the local curvature at any point, thus separating space into
distinct regions depending on the sign of these eigenvalues
$\lambda_i$. Within a 3D space, as by definition $\lambda_j<\lambda_i$
if $j>i$, there exist four of these regions. Let $I$ be the number of
negative eigenvalues, then the regions where $I$ is equal to 0, 1, 2
and 3. This
classification applies to critical points of the field in particular,
where $\nabla \rho=0$, the maxima ($I=3$) and minima ($I=0$) existing
within local clumps and voids respectively, while two types of saddle points
can be distinguished: the filaments type saddle points (for $I=2$) and
the pancake type ones (for $I=1$).\\

Figure \ref{fig_courbure3D} illustrates a second order approximation
of the density field in the vicinity of the field extrema. The total
set of critical lines form a fully connected path linking all the
critical points together and exactly six branches pass through each of them in the direction of  the three
eigenvectors of the Hessian. 
Empirically, it is possible to picture the typical behavior of the whole set of critical lines.
Defining $E = \lbrace
0,1,2,3\rbrace$ and considering a given critical point where $I=n$,
 if $i<j<k\in E-\lbrace n\rbrace$, this critical point $C_n$ is usually
linked to three other pairs of  critical points $C_i$, $C_j$ and $C_k$ (where
$I=i$, $I=j$ and $I=k$ respectively)  by critical lines aligned with eigenvectors associated to eigenvalues
$\lambda_1$, $\lambda_2$ and $\lambda_3$ respectively at point
$C_n$. Most of the time, each of these branches connect to critical point $C_i$, $C_j$
and $C_k$  along the eigenvector associated with eigenvalue
$\lambda_1$, $\lambda_2$ and $\lambda_3$ respectively, evaluated at
points $C_i$, $C_j$ and $C_k$ respectively. In this picture,
the critical lines can be seen as a fully connected path
linking all the different regions defined by the sign of the
eigenvalues of $\vH$.\\

The overdense filamentary structure
correspond to the subset of the critical lines that constitute an
approximation of the ``real'' skeleton (i.e. the ``ridges'' of the
distribution). This part is the one which links maxima ($I=3$) and
filamentary saddle points ($I=2$).  The typical behaviour of such
lines is the following: In the immediate vicinity of a non-degenerate
maximum, two branches of the skeleton exist, stretching in the
eigendirection that corresponds to $\lambda_1$. Following one of
the branches, denoting as $\lambda_{||}$ an eigenvalue whose eigenvector is parallel to the skeleton and $\lambda_{\perp,1,2}$ as two
eigenvalues associated to eigenvectors in the perpendicular directions.
 Near the maximum, $0 >
\lambda_{||}=\lambda_1 > \lambda_{\perp,1} > \lambda_{\perp,2}$.  As
one follows a branch one probable outcome is the change of sign of
$\lambda_{||}$, in which case the branch will typically end in a saddle
point of a filamentary type along its $\lambda_1$ direction. There is
always another branch that starts from this saddle point on the other side,
thus this type of branches have a fully connected structure.  However,
another possible outcome is that one of the orthogonal eigenvalues
changes faster than $\lambda_{||}$ as one moves away from the maximum
and becomes positive before the saddle point is reached. In this case
the branch of the local skeleton formally terminates, which however in
reality often means that the skeleton splits at this point in two new
branches.\\

Such branching of the skeleton is especially frequent near the maxima
of the field, where it accounts for how multiple filamentary sections can
end up in a single dark matter halo. Studiing how skeleton segments merge is
relevant for questions such as the multipole structure of matter inflow
onto dark halos (Aubert, Pichon \& Colombi 2004, Pichon \& Aubert 2006). 
This property of skeleton segments to end outside of the critical
points is specific to the local definition of the skeleton, in contrast to
 the ``real'' skeleton whose segments are always
connected on both ends.

\section{The skeleton length for scale-free Gaussian random fields}
\label{sec:gauss}
Before considering general cosmological density fields, 
the local skeleton of  scale free Gaussian random fields $\rho$
with null average value $\langle\rho\rangle=0$ will be investigated. For convenience, it is useful to
define some spectral parameters that depend on the spectral index $n$ and on the smoothing length.
In the statistical description of the skeleton of a random density field (Appendix C), the following spectral parameters appear to play a role:
\begin{eqnarray}
\displaystyle \sigma^2_0 & =& \langle\rho^2\rangle,\\ \displaystyle
\sigma^2_1 & =&  \langle\left(\nabla \rho\right)^2\rangle,\\ \displaystyle 
\displaystyle \sigma^2_2 & = &  \langle\left(\Delta \rho\right)^2\rangle,\\ \displaystyle 
\displaystyle \sigma^2_3 & = & \langle\left(\nabla \Delta \rho\right)^2\rangle.
\label{eq:sigmadef}
\end{eqnarray}
This introduces three linear scales into the skeleton theory
\begin{equation}
R_{0} = \frac{\sigma_{0}}{\sigma_{1}}, \quad R_{*} = \frac{\sigma_{1}}{\sigma_{2}}, \quad \tilde R  = \frac{\sigma_{2}}{\sigma_{3}}
\label{eq:rdef}
\end{equation}
where the first two have a well-known meaning of typical separation between zero-crossing of the field $R_{0}$ and  mean distance between extrema, $R_{*}$ (Bardeen \& al. 1986) ,
while the third one, $\tilde R$ is, by analogy, the typical distance between the inflection points.

Out of three scales two dimensionless ratios may be   constructed  that are intrinsic parameters of the theory 
\begin{equation}
\displaystyle \gamma   \equiv \frac{R_{*}}{R_{0}}=\frac{\sigma_1^2}{\sigma_0\sigma_2} , \quad
{\tilde \gamma}  \equiv \frac{\tilde R}{R_{*}}=\frac{\sigma^2_2}{\sigma_3 \sigma_1} \,,
\label{eq:gammadef} 
\end{equation}
where $\gamma$ says
 how frequent encountering a maximum between two zero crossings of the field is, while $\tilde \gamma$ describes, on average, how many inflection points are between two extrema.
For  Gaussian fields, these parameters can be easily calculated from the power spectrum. Both  $\gamma$ and $\tilde \gamma$ range from zero up to one.
For reference, for the power-law spectra with index $n > -3$, smoothed at small scales with a Gaussian window,
\begin{equation}
\gamma = \sqrt{\frac{n+3}{n+5}}, \quad \tilde\gamma = \sqrt{\frac{n+5}{n+7}}\,. \label{eq:thegamm}
\end{equation}
Note that  cosmologically relevant density power spectra have $n > -3$ and thus, while $\gamma$ can attain low values, $\tilde\gamma$ are always close to unity\footnote{Cosmological density fields, therefore, have of order one inflection point per extremum, unlike, for, example, a mountain range, where one encounters many inflection points on a way from a mountain top to the bottom} .  

Appendix~C introduces a statistical description of the skeleton for  the Gaussian and non Gaussian random field.
This section presents the numerical measurements of the properties of the skeleton for scale free Gaussian fields.

The first quantity of interest is the total length of the skeleton,
$L_{\rm tot}$. 
 In the context of cosmology, $L_{\rm tot}$ can
be linked to the total length of the filaments linking clusters
together and in that sense reflects the history of matter accretion as
well as the initial distribution of matter (which is supposed to be
similar to a Gaussian random field with a scale-dependent
 effective spectral index similar to the
ones considered here). Figure \ref{fig:len} presents the result of
the  measurement of the total length $L_{\rm tot}$ of the skeleton per
unit box size as a  function of the spectral index and for different
smoothing lengths $\sigma$  (within the range of validity of the algorithm 
as described in Appendix B).  These measurements are carried 
  over $25$ realisations of scale free  $256^{3}$ Gaussian random fields as a function of
  the spectral index $n$.
 The sensitivity of the skeleton to the
value of the spectral index is  clear on this plot and, if
$L_{\rm tot}$ appears to be a linear function of the spectral index,
it is also clear that it grows as a power law of the smoothing
length. The dotted lines on figure \ref{fig:len} shows the result of
such a fit of the data and seems to work very well. A very good
approximation of $L_{\rm tot}$ per unit box size is thus given by the
function:
\begin{equation}
L_{\rm tot} = 0.21(n+5.28)\sigma^{-2.00}.\label{eq:defLL}
\end{equation}

As expected, the exponent of $\sigma$ is measured to
be exactly $2$. It can be proved with a simple argument that this
should be the case for scale free Gaussian fields. In fact, for such
fields, computing the skeleton over a grid of volume $l^3$ and
smoothed on a scale $\sigma$ is equivalent to computing the skeleton
on a grid of volume $\left( \alpha l\right)^3$ while smoothing on a
scale $\left(\alpha\sigma\right)$ and rescaling the result by a factor
$1/\alpha$. Because of the scale invariance, we also have
$L\left(\sigma\right) = \alpha^{-3} L\left(\alpha\sigma\right)$ and so
$L\left(\sigma\right)\propto \sigma/\sigma^{3} = \sigma^{-2}$.

%

\begin{figure}
  \centering \includegraphics[angle=0,width=8cm]{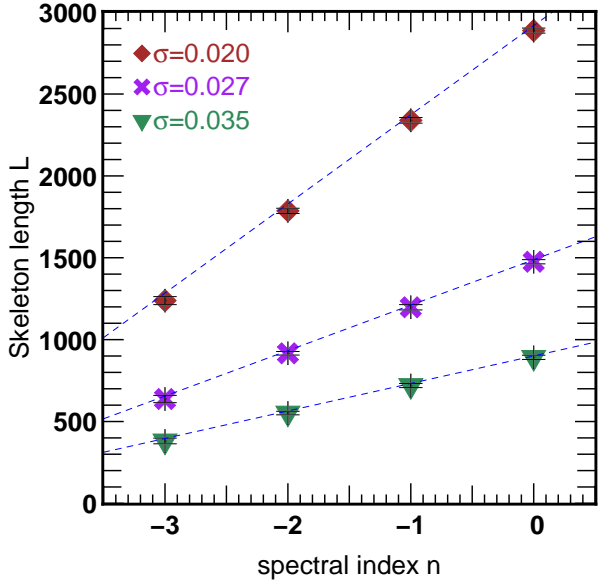}
  \caption{ Total length $L$ of the skeleton per unit box size for
  different smoothing lengths $\sigma=0.020,0.027,0.035$; measured
  over $25$ realisations of Gaussian random fields as a function of
  the spectral index $n$. While $L$ depends linearly on the spectral
  index $n$, it grows as a power of $\sigma$. The dotted lines
  represent the fits obtained  using the function: $L \approx
0.21  (n+5.28)\sigma^{-2}$. \label{fig:len}}
\end{figure}
Interestingly, the dependence on the spectral index $n$ is close to
$n+5$ which argues for filaments being relatively straight between 
extrema, see Appendix~C. 
A visual examination of the filaments confirms this picture. 

Now consider the differential length of the skeleton,
$dL/d\eta\left(\eta\right)$ where $\eta \equiv\rho/\sigma_0$ is the normalized density contrast. This quantity represents the expected
length of skeleton  that can be measured in a given distribution
between density contrasts $\eta$ and $\eta+d\eta$.
Figure
\ref{fig:diflen} shows the normalized function
$dL/d\eta\left(\eta\right)$ as a function of the normalized density
contrast $\eta$ from which was subtracted  the probability distribution
function (PDF) of the field (which, within the range of sampling and finite volume effects approximations, is a Gaussian function). 
These values were also averaged over $25$
realisations of Gaussian fields with spectral index $n=0,-1,-2$
sampled on $256^3$ pixel grids and for a smoothing length
$\sigma=0.027$. This value was chosen as a compromise between finite 
volume effect and differentiability of the field on a grid
discussed  in Appendix~\ref{chap:slen}. 
Considering the error bars, it is clear that the value of
$dL/d\eta\left(\eta\right)$ is directly linked to the spectral index
$n$.


It is shown in Appendix~C that 
$dL/d\eta\left(\eta\right)$ can be written 
using an Edgeworth expansion (see also Novikov, Colombi \& Dore 2005  for the corresponding proof and  fit 
in 2D):
\begin{equation}
\frac{dL}{d\eta}\left(\eta\right)=\frac{L_{\rm
tot}}{2\pi}\exp\left({-\eta^2/2}\right)\left(\sum_{n\ge 0}
C_{2n}\gamma^{2n}H_{2n}\left(\eta/\sqrt{2}\right)\right),\label{eq:fitdldn}
\end{equation}
where $L_{\rm tot}$ is the total length of the skeleton, $C_0=1$ and
$H_{2n}$ are Hermitte polynomials. 
Figure~4 demonstrates that this expansion also
works very well in the 3D case.  Remarkably, equation~(\ref{eq:fitdldn}) does not depend on $\tilde \gamma$ which
again argues for the picture of a  stiff  behaviour of the skeleton for cosmological scale invariant density fields (see Appendix~C).
 Table \ref{tab1} presents the
values of the first three coefficients $C_{2n}$ obtained by fitting
the measurements presented in figure \ref{fig:diflen} (the dotted line
of figure \ref{fig:diflen} are the result of these fits). Not only
does equation~(\ref{eq:fitdldn}) allows a very good fit of the
measured data, but it also appears that only the first order term is
non-null and the differential length of the skeleton of a Gaussian
random field with spectral parameter $\gamma$ is thus given by:

\begin{equation}
\frac{dL}{d\eta}\left(\eta\right)=\frac{L_{\rm tot}}{2\pi}\exp\left({-\eta^2/2}\right)\left(1+ 0.21 \gamma^2\left(\eta^2-1\right)\right)\label{eq:fitdldn_val}.
\end{equation}
The only non-null coefficients in the expansion are thus $C_0=1$ and $C_2=0.21$, to be contrasted to $C_2=0.17$ in the 2D case.
Equation~(\ref{eq:fitdldn_val}) can be used as a test of non gaussianity like any other topological estimator, such as the genus, the PDF etc... as discussed 
in Novikov et al.(2006), since departure from the shape of equation~(\ref{eq:fitdldn_val}) must appear when the skeleton's differential length is 
computed while  the underlying field is not Gaussian\footnote{of course, given the properties of the skeleton,
 this won't apply if the non Gaussianity involves only a (monotonic) bias}. 
\begin{figure}
  \centering \includegraphics[angle=0,width=8.5cm]{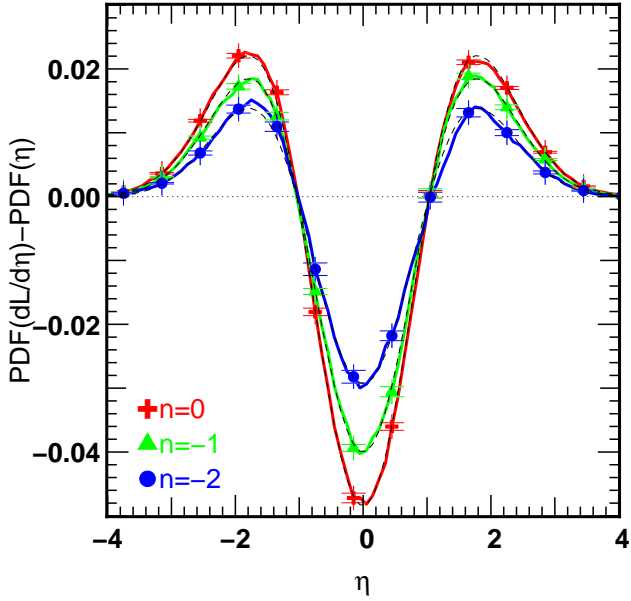}
  \caption{Difference between the probability distribution function
(PDF) of the density field and the normalized differential length of
the skeleton $dL/d\eta$ as a function of the density contrast
$\eta=\rho/\sigma_0$. Each curve represents the average value and
variance of the measured value of $dL/d\eta$ over $25$ different
realisations of scale-free Gaussian fields, for different values of
the spectral index $n=0,-1,-2$. The dotted curves represent the
estimation obtained by fitting data using equation~(\ref{eq:fitdldn})
(see table \ref{tab1} for values of the parameters). \label{fig:diflen}}
\end{figure}

\begin{table}
\begin{tabular}{|c|c|c|c|}
\hline & $C_2$ & $C_4$ & $C_6$\\  \hline $n=0$ &$0.219$& $0.006$&
$-0.001$\\ $n=-1$ &$0.212$& $0.002$& $-0.002$\\  $n=-2$ &$0.206$&
$-0.005$& $-0.008$\\  \hline &$0.21\pm0.005$& $0.001\pm0.005$&
$-0.004\pm0.003$\\ \hline
\end{tabular}
\caption{Measured values of the first three non-null terms in the
Edgeworth expansion, equation~(\ref{eq:fitdldn}), for three different values of
the spectral index $n=0,-1,-2$. These results are obtained by fitting
equation~(\ref{eq:fitdldn}) on the data presented in figure
\ref{fig:diflen} on which the  dotted lines represent the fitted
function. The measurements show very good agreement, whatever the
value of $n$.\label{tab1}}
\end{table}

 For the matter distribution in
the universe, the filaments are overdense regions along which matter
flows. In that sense, they are less subject to numerical or
observational noise and contain most of the information about the underlying matter
distribution.
 The skeleton length can thus be seen as a method for
measuring  the power spectrum which naturally weights information in different
regions  according to their importance.

\section{Illustration:  dynamical environment of filaments}

Drawing the skeleton allows us to pin down the nature of the flow
around the filaments. Indeed one may roughtly define three dynamically
distinct regions in large-scale structures: voids, clusters and
filaments. The first two have been investigated in some detail.  The
filaments represent a fairly unexplored venue. Beyond  the kinematics
(velocity distribution,  spin, etc.), the photometric and
spectroscopic properties of galaxies (colour, age, metallicity etc..),
their morphology (ellipticals versus spirals, Gini number, Asymmetry)
or the IGM (gas temperature, WHIM detection, fraction of gas/metals in
the filaments etc..), could also be investigated as a function of the
distance to, and along the filaments.

In this section, two examples simply illustrate how the skeleton can
be used  to explore the environment of filaments in cosmological
simulations.

\subsection{ Dark matter flow near  the skeleton }

\begin{figure*}
\centering
\subfigure[]{\includegraphics[angle=0,width=8.5cm]{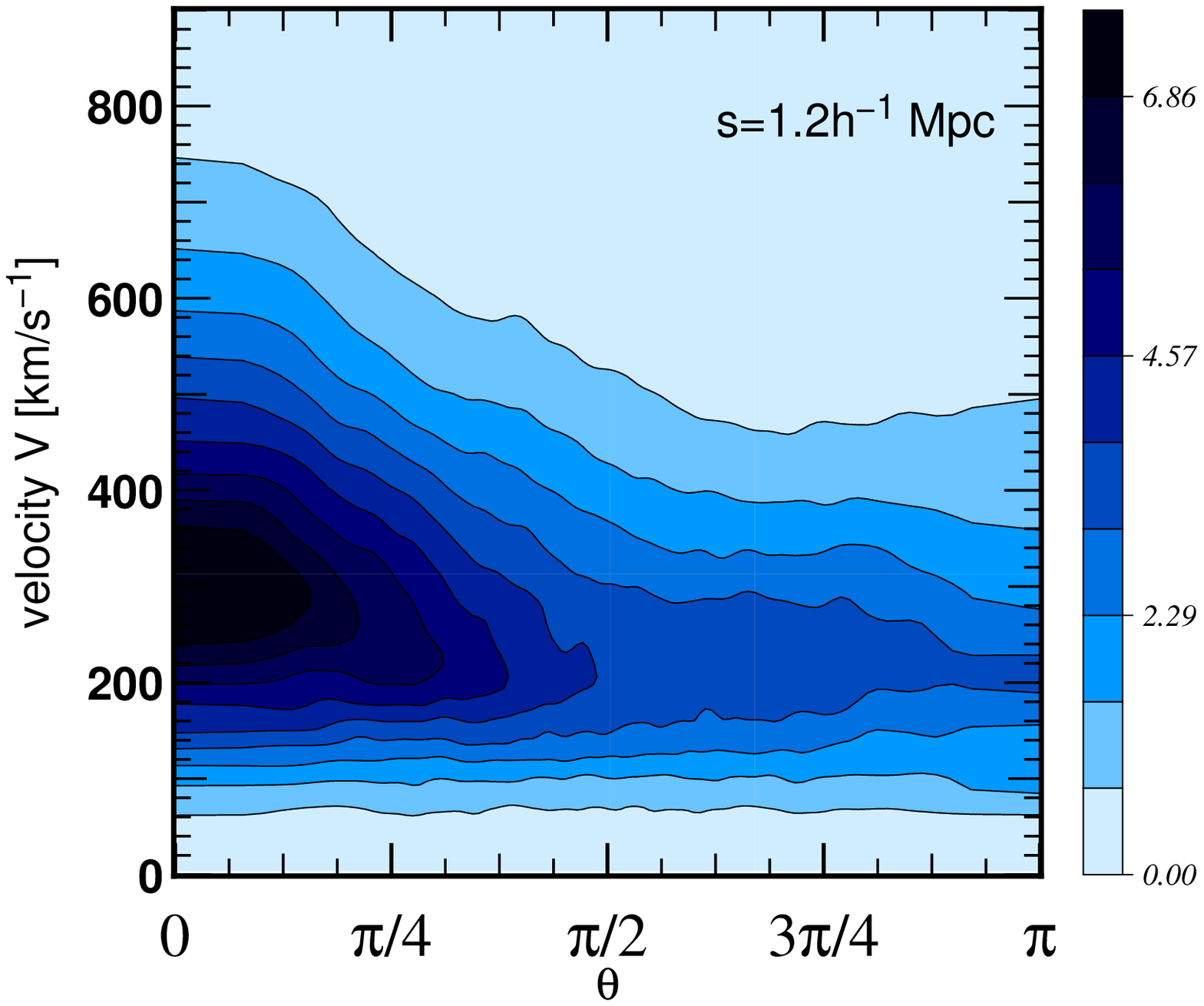}\label{fig:vel1p2}}
\hfill
\subfigure[]{\includegraphics[angle=0,width=8.5cm]{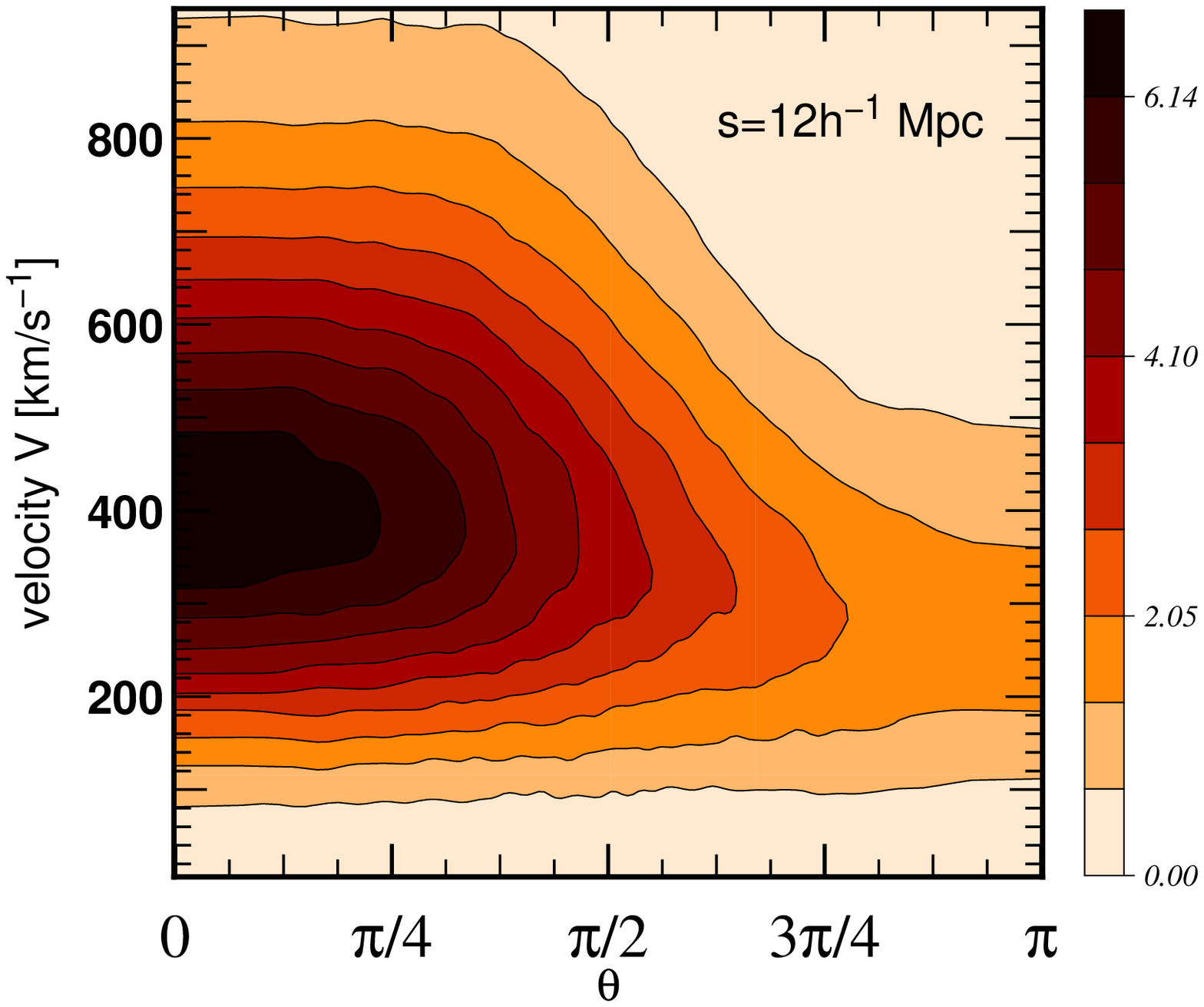}\label{fig:vel12}}\\
\centering
\subfigure[]{\includegraphics[angle=0,width=8.5cm]{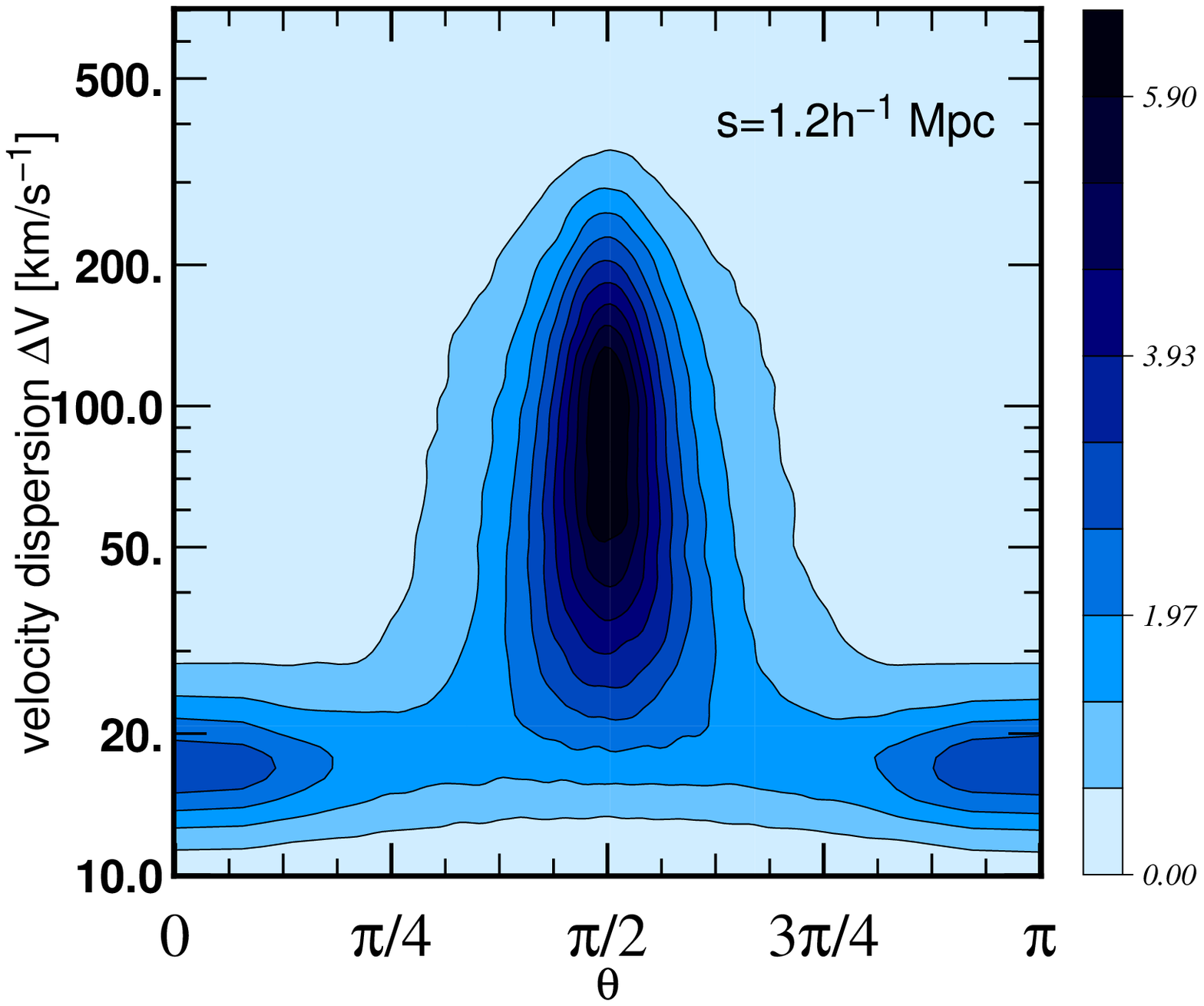}\label{fig:disp1p2}}
\hfill
\subfigure[]{\includegraphics[angle=0,width=8.5cm]{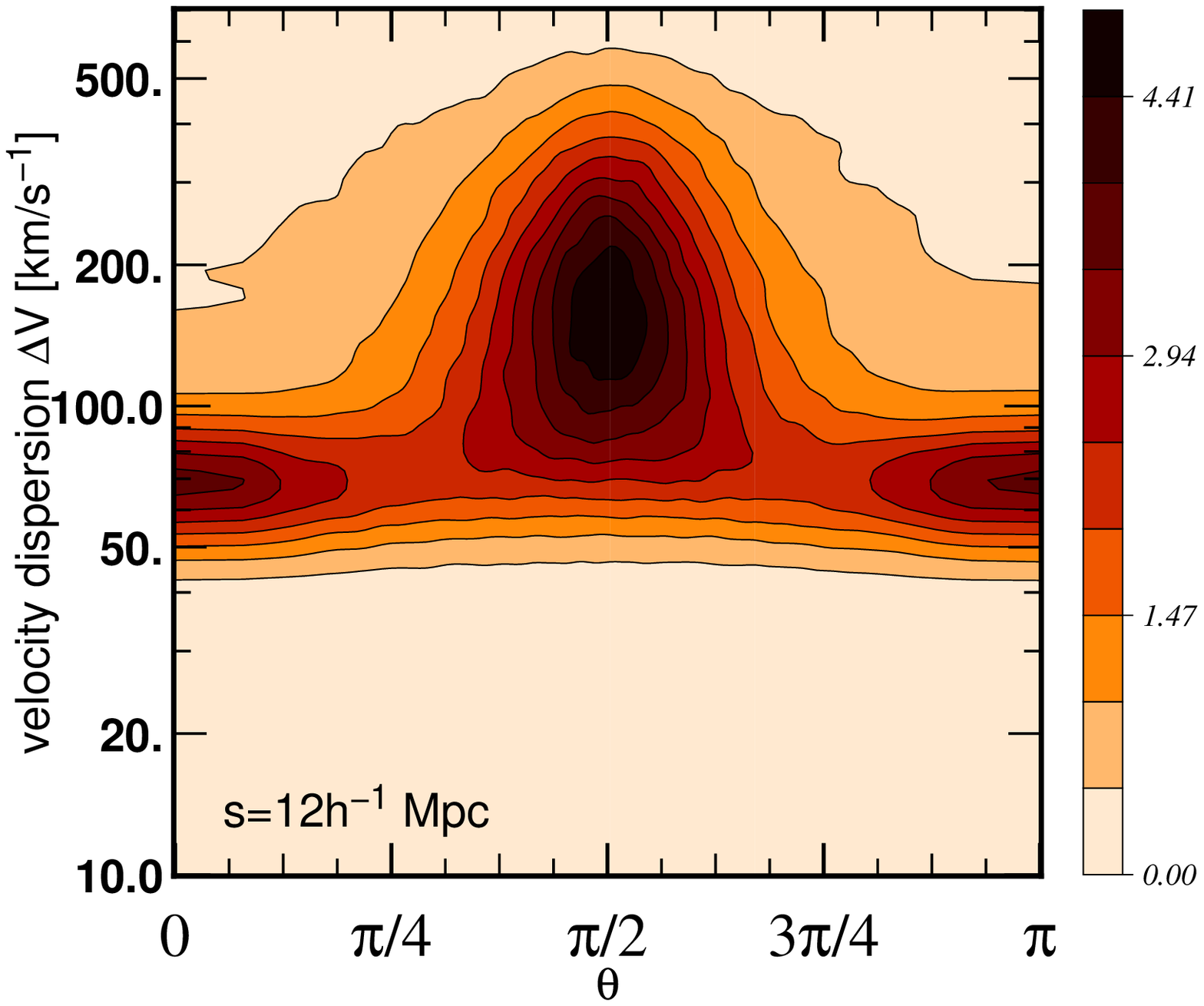}\label{fig:disp12}}
  \caption{ {\it Top panels:} Probability distribution function (PDF)
    of the velocity field $V$ of the dark matter along the skeleton
    as a function of its angle $\theta$ with the skeleton and its
    norm. The measurements were achieved on a $100h^{-1}$ Mpc and
    $1000h^{-1}$ Mpc dark matter  simulation featuring $512^3$
    particles and a standard $\Lambda$CDM model, smoothed over  a
    scale $s=1.2h^{-1}$ Mpc  and $s=12h^{-1}$ Mpc ({\it left} and {\it
    right} panels respectively). The skeleton is oriented in the
    direction of increasing density. Dark matter appears to be flowing along the
    filaments in the direction of higher density regions
    (i.e. halos). {\it Bottom panels:} PDF of main eigenvector of the
    velocity dispersion tensor $\Delta V_{ij}$ as a function of its angle
    $\theta$ with the skeleton and its eigenvalue amplitude. The peak of the PDF corresponds to 
    high velocity dispersion orthogonal to the filaments, which is
    coherent with the picture of dark matter being accreted
    orthogonally by the filaments before flowing along them. Note the
    increase in velocity dispersion with scale ({\it left} and {\it right panels}) as well as the larger
    angular dispersion in the dark matter flow.
    This trend is also found 
    while considering the same simulation at higher z.
\label{fig:PDF}}
\end{figure*}

Figure \ref{fig:PDF} displays probability distribution functions (PDF)
of different characteristics of the dark matter flow along the
skeleton. In order to understand the correlations between the
filaments and the velocity field, we computed the PDF of its angle relative to
 the skeleton as a function of its intensity (top panels), and 
 the PDF of the angle between its largest eigenvector and the
skeleton as a function of the norm of the corresponding
eigenvalue (bottom panels). These measurements were achieved by first sampling the
field characteristics on a grid, averaging particles velocities $V\equiv\langle v\rangle$
and dispersion tensor, $\Delta V_{ij}^{2}\equiv\langle(v_{i}-\langle v_{i}\rangle)(v_{j}-\langle v_{j}\rangle)\rangle$ over each cell, and then
computing for each segment the distance-weighted average of their
PDF.
Left and right panels yield the resulting PDF computed in a
$100h^{-1}$ and $1000h^{-1}$ Mpc dark matter standard $\Lambda$CDM
model simulation respectively, at redshift $z=0$ and using $512^3$ particles. In both cases, the density and velocity fields where
sampled on a $512^3$ pixels grid and smoothed over $\sigma_p=6$ pixels
(i.e. $s=1.2h^{-1}$ Mpc and $s=12h^{-1}$ Mpc respectively). The
skeleton segments being oriented in the direction of increasing
density, an angle of $\theta=0$ means that dark matter is flowing
along the filament in the direction of higher density regions.\\

The flows appears to be laminar and its amplitude increases with
scale: this is expected since on larger scales the clusters are more
massive, the potential difference is larger, hence the flow towards
them is faster.  Most dark matter particles have a mean velocity of
about $300$ (resp. 400) km/s along the filament and a dispersion of
about 100 (resp. 150) km/s orthogonal to the filaments for the two
scales considered here.  The angular spread
(panels~\ref{fig:vel1p2}-\ref{fig:vel12}) also increases with scale,
from about $30^{\rm o}$ to about $45^{\rm o}$, reflecting the larger
internal heat of the filament, also seen in
(panels~\ref{fig:disp1p2}-\ref{fig:disp12}).  \\

  The qualitative shape of
this PDF may be explained by the advection of new halos onto the
``highways'' corresponding to the mean flow.  The first eigenvector of the
dispersion tensor is on average clearly orthogonal to the filament, reflecting
the velocity of dark matter falling onto the filaments.  Note that the
distribution is decreasing monotonously with $\theta$ in
panel~\ref{fig:vel1p2}: some dark matter particles statistically even
move downhill, and their relative fraction decreases with scale. The
filaments are collecting matter away from the underdense regions.
Smaller filaments empty smaller voids, which tend to get depleted
earlier than larger ones; hence this may explain why the flow becomes
more orderly at smaller scale as accretion diminishes.
\\
Note that the redshift evolution (not shown here)
 of this distribution follows closely its scale evolution,
the $z=15$ PDF over $100 h^{-1}$ Mpc resembling the $z=0$ PDF over $1000 h^{-1}$ Mpc \cite{TheseSousbie}.
\\

The detailed nature of the flow should eventually be investigated 
in a smoothing scale independent manner, in order  to derive 
universal features which would only depend on the cosmology 
and the initial power spectrum. Its evolution with redshift or with the cosmology should also 
be systematically analysed.

\subsection{ Dark matter spin-skeleton connection }

The geometric orientation of the spin of dark matter halos corresponds
 to another feature of the large scale structure which can be
 characterized using the skeleton.  The spin of dark halos was
 computed using the classical friend-of-friend (FOF) algorithm  with  0.2 times the interparticular distance 
as linking length and retaining only halos containing more than 100 particles.
 Figure~\ref{fig:spin} displays the excess probability of alignment of
 the halos' spins with the closest skeleton segment for different distances 
 $ [0,0.5]$, $[0.5,1.5]$, $[1.5,2.5]$ and $ [2.5,3.5] h^{-1}$ Mpc.  This probability
 reaches 25 \% for an angle $\theta=\pi/2$ between the spin and the
 skeleton: the spin of dark matter halos is preferentially orthogonal
 to the filament they belong to.  This trend accounts for the fact
 that the filaments are the locus of laminar flow where halos coalesce
 along the direction of the filaments parallel to the mean flow, hence
 acquiring momentum orthogonal to the flow, as observed in (Aubert, Pichon \& Colombi 2004).

\begin{figure}
  \centering \includegraphics[angle=0,width=8cm]{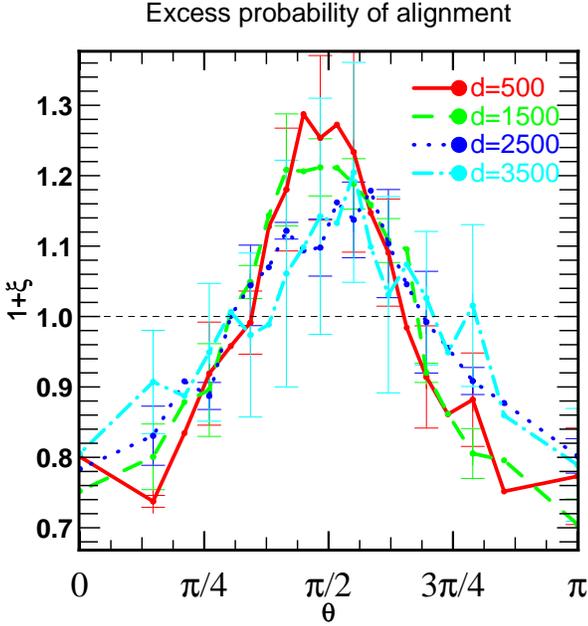}
 \caption{ Excess probability of spin alignment with the local
 skeleton computed from the average of three $512^{3 }$ $50 h^{-1} $
 Mpc $\Lambda$CDM simulations at different distances: $d \in [0,500],
 [500,1500],[1500,2500]$ and $ [2500,3500] h^{-1}$ kpc from the
 closest skeleton segments. This figure demonstrates that on average
 the spin of dark matter halos tends to be orthogonal to the local
 filaments at a level of 25 \% for distances shorter than 500 kpc.  The
 simulation is analysed at redshift zero.
   \label{fig:spin}
   }
\end{figure}

\section{Conclusion \& perspectives}

The 3D skeleton formalism is a well-defined framework for studying
the filamentary structure  of a distribution. The``real'' skeleton is
defined as the set of critical lines joining saddle points to maxima
of the field along the gradient. A local approximation of it was
introduced in section \ref{sec:3dskl} along with a numerical method
allowing a fast retrieval of the locus of the filaments from a sampled
field (see also appendix \ref{imple}). This method involves computing  the null isodensity surfaces
of each component of a function ${\cal S}=\, (\,\vH\cdot \nabla
\rho \times \nabla \rho\,)$ of the gradient, $\nabla \rho$, and Hessian matrix,
$\vH$ of this field.\\

The ability to localize and characterize the filamentary structure of
matter distribution in the universe opens the prospect of 
many applications for the skeleton as discussed in Section~4 and 5. It has been shown in section
\ref{sec:gauss} that for a  Gaussian random field, the total length of the skeleton per unit volume
depended only of the chosen smoothing length $\sigma$ and spectral
index $n$, with a specific functional from which was both fitted from simulations 
and motivated in Appendix~C. 
In this sense, the local skeleton
provides a direct measurement of the local shape of the power spectrum,
$P\left(k\right)$, on various scales depending on the smoothing applied to the underlying
field. Though there exist other ways to measure
the power spectrum of a given distribution, the skeleton length
is promising as it relies only on the filamentary
structure of the distribution. 
The analysis of the length of
the skeleton of the galaxy distribution in the SDSS as a measurement
of cosmological parameter $\Omega_m$  can be found in \cite{SouSkLet}.\\

The skeleton may also be used as an isotropy probe.  It corresponds in fact to 
a good candidate for the Alcock-Paczynski \cite{Alcock} test, since the apparent 
longitudinal to transverse length of skeleton segments should directly 
constrain the curvature of space in a manner which is bias-independent.
This test will be presented in a forthcoming letter \cite{letterAPtest}.
\\

It was demonstrated in section~4 that the dark matter flow in the
vicinity of filaments  was dominantly  laminar along the filaments and
shows signs of orthogonal accretion corresponding to  the infall of
dark matter collected from the voids. It also showed that the spin of
dark matter halos were preferentially orthogonal to the filament's
direction, a feature which can be understood as a consequence  of
merger events taking place along these filaments.  A clear virtue of
the local skeleton is that since it relies on a local expansion of the
field, it can deal with truncated/masked fields, segmented  or
vanishing ridges or isolated structures. Note finally that the fit,
equation \ref{eq:fitdldn_val}, opens the prospect of using the local
skeleton  to estimate the bias in observed surveys.  The idea is to
compute the PDF of galaxies on the one hand, which  depends on the
mass to  light ratio of the sample, and the differential length
(equation~(\ref{eq:fitdldn_val})) on the other hand.  Since the former
depends on the bias, whereas the later does not, comparing the two
should give an estimate of the bias.  On the other hand, the local
formulation of the skeleton presents some limitations. Mainly, it is
not fully connected: it has by construction (since it is drawn from a
second order Taylor expansion of  the field) only 2 segments per
maxima whereas full connection would require 3 or more. A consequence
is that it cannot represent merging filaments. \\ 

One could also use the
curvature and torsion of filaments as cosmological probes,  since the
acceleration of the universe induced by the cosmological constant is
likely to straighten the filaments, though the fact that the local
skeleton has only two segments near its maxima (the other segments
must branch out) is likely to introduce some artifacts.  The topology
and geometry of the skeleton near the density peaks and the redshift
evolution of the skeleton of the large scale structures may prove of
interest, for instance to study the frequency of reconnection, though
again the local skeleton is not ideal  in this respect.  It would also
be interesting to construct the skeleton in higher dimensions,  for
instance in space-time, to trace the events lines, but again
connection is critical.  In a forthcoming paper, an alternative
algorithm for the indentification of the skeleton,  loosely based on a
least action formulation, will be presented. It is  complementary to
the solution presented in this paper and will allow us to tackle those
points for  which the local skeleton is less efficient.  Finally, the
3D skeleton algorithm  could possibly be applied to  other fields of
research, such as neurology, in order to trace the neural network.

\section*{Acknowledgments}
We thank H\'el\`ene Courtois, D.~Aubert and Simon Prunet for comments   and D.~Munro  for  freely
   distributing   his  Yorick  programming   language and opengl interface  (available   at  {\em\tt
  http://yorick.sourceforge.net/}).
This  work  was  carried  within  the framework of the   Horizon  project,
\texttt{www.projet-horizon.fr}.


\appendix

\section{Numerical implementation}
\label{imple}
All the computations were performed using a specially developed \texttt{C}
package: {\bf \texttt{SkelEx}}\footnote{Available on request from the authors.}
({\bf Skel}eton {\bf Ex}tractor). This package also includes a flexible
OpenGL visualization tool that was used for making the figures in this paper.\\

The first step before computing the skeleton requires obtaining a
density field from a discrete point-like distribution.  This is
achieved by smoothing appropriately the density field on a grid so
that it is not singular (i.e. is sufficiently differentiable) but
still contains all the topological information. The density field is
computed using Cloud-In-Cell (CIC) interpolation (e.g., R.W. Hockney
1988) and convolving the result with Gaussian windows of different
widths. As was shown in section \ref{chap:slen}, the grid size and
smoothing length are decisive parameters. It is then necessary to
compute first and second derivatives of the field on the grid, which
can be done using finite difference or Fourier transform method,
giving very similar results. In fact choosing one method or the other
does not seem to have any influence on the resulting skeleton if the
field is smooth enough (which is anyway a necessary condition).\\

The next step  involves solving the
system of equations~(\ref{eq:3dS_0}); the solution of this system
corresponding to the intersection of two of the three solutions of
equations~(\ref{eq:3dSi}). This is done by computing the 3D  meshes of
the two-dimensional surfaces that are solution to these equations: the
skeleton is at the intersection of two of them, depending on the value
of the gradient at the point considered. Solving equation
~(\ref{eq:3dSi}) is equivalent to finding the null isocontour of
field ${\cal{S}}_i$, which can be done using the marching cube
algorithm \cite{marchingcube}. The basic idea is to consider every cell of the grid as an
individual cube. One can then compute the value of every ${\cal{S}}_i$
for the eight vertices and it is easy to check whether the
isosurface intersects the cube or not. In fact, every vertex is above
or below a threshold value (in this case $0$), which gives a total of
$2^8=256$ types of intersections (only 15 of them being instrinsically different)
that can be precomputed as illustrated in figure
\ref{fig:mc_config}. The exact positions of the intersections are
computed using quadratic interpolation. This yields the position of the intersections of the grid and the
isocontour, and defines triangles that smartly link those intersection
vertices:  one can  then reconstruct a very good approximation of what the
isocontour is.\\

\begin{figure}
  \centering \includegraphics[angle=0,width=8cm]{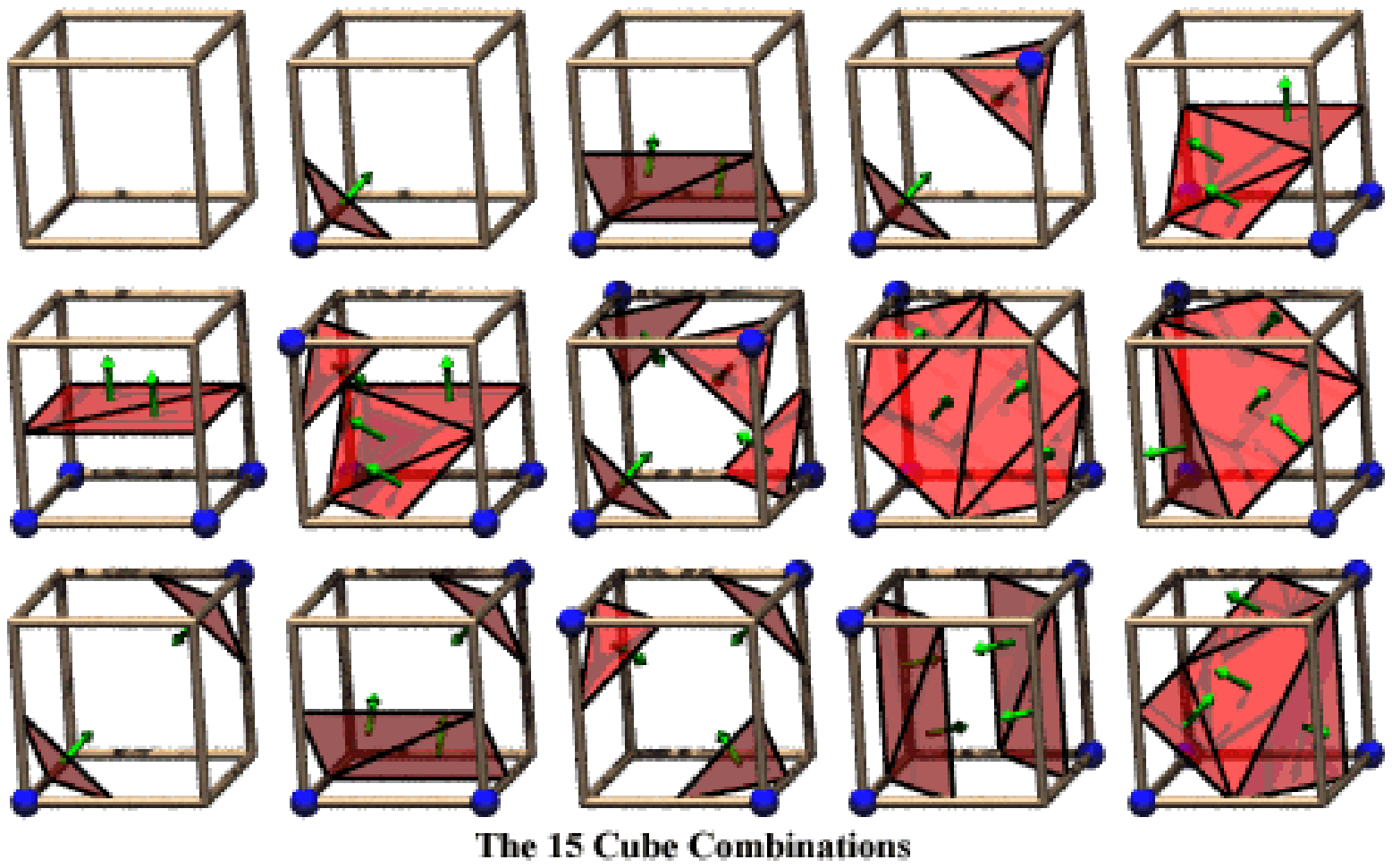}
  \caption{ Illustration of the different possible configurations of a
    grid cell used for marching cube algorithm. Given a field $f$ and
    isocontour $f=0$, a blue ball represents a vertex where $f>0$. It is then
    easy to build the isocontour by linearly interpolating the value of $f$
    along the edges. This picture was borrowed from James Sharman's website,
    http://www.exaflop.org/docs/marchcubes/ind.html }
\label{fig:mc_config}
\end{figure}

Which surfaces should be used for each cell is decided by computing
$d_k={\rm det}\left(\bf{r_i},\bf{r_j},\bf{\nabla\rho}\right),\; i \neq j
\neq k \in \lbrace 1,2,3\rbrace$ and selecting only the two
${\cal{S}}_k$ for which $d_k$ is maximal. This gives two surfaces defined
by triangles whose intersection can be efficiently computed:  it amounts
 to computing the intersection of triangle pairs only. It is
then straightforward to compute the eigenvalue of the Hessian for every segment
and keep or reject  them depending on the previously defined criteria
(equation~(\ref{eq:eigselect3D})) in order to draw the {\it{local}}
skeleton. The exact same method was used for efficiently  and consistently finding the
extrema and saddle points of the field. In fact, if one defines three
fields $f_i={\partial \rho}/{\partial r_i}$, those critical points
are the intersections of the three isocontour surfaces $f_i=0$. One
can then decide if a critical point is a maximum, minimum or saddle
point by checking the value of eigenvalues of the Hessian
(i.e. curvature).
Although marching cubes algorithm are very efficient for computing
isodensity contour, they present some drawbacks for ambiguous configurations. Indeed, as illustrated on
figure \ref{fig:singular}, some configurations are degenerate and one
cannot decide where the isosurface should pass. This problem happens
most of the time around critical points where the value of the field
can go above and below the threshold within one cell. It induces
the loss a small skeleton segments.

\begin{figure}
  \centering \includegraphics[angle=0,width=8cm]{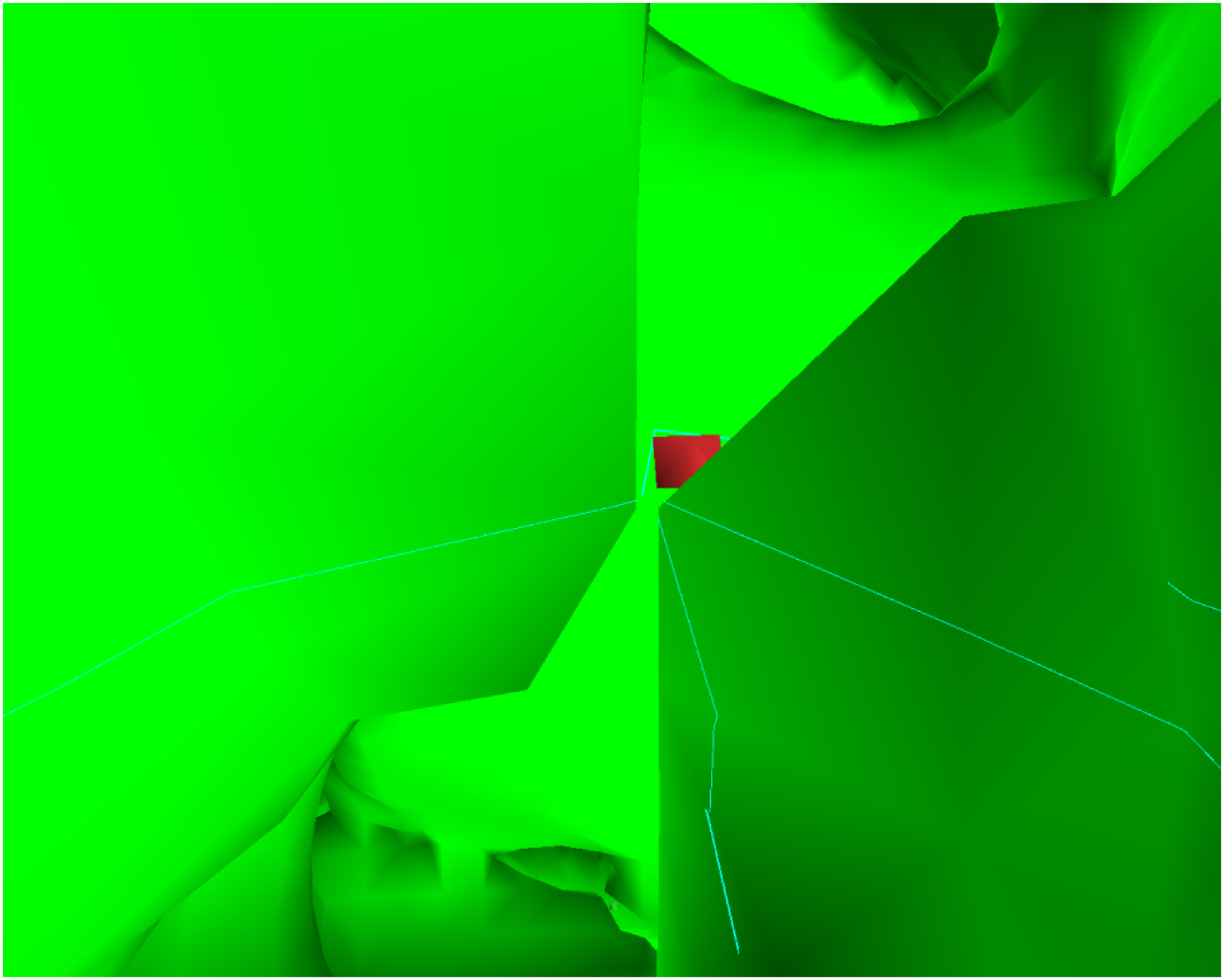}
  \caption{ Illustration of a drawback of marching cubes algorithm. The
    green surface is an isosurface solution of equation~(\ref{eq:3dSi}) and the
    light blue line is the resulting skeleton. The red diamond represent a
    field maximum. It is clear on this picture that the algorithm misses the
    part close to that maximum, thus creating a spurious hole in the
    skeleton.}
\label{fig:singular}
\end{figure}

In order to obtain a smooth skeleton that does not present holes and
to retrieve connectivity information (i.e. to be able to follow the
skeleton from one point to another), a three steps post-processing is
applied. There is of course some arbitrariness in this process:
here  the
algorithm is based on a weighted marking system to achieve this result (where the weights are assigned depending
on the relative importance of the selection criteria):
(i)  the  branches that were missed around the
extrema are regenerated using the fact that the skeleton around an
extremum is along  the main curvature axis (i.e. along the first eigenvector of the Hessian). 
So for each extremum, marks are given to all skeleton segment, favoring those at small
distances and with similar orientation as the main eigenvector of $\vH$.
Each extremum is eventually connected to the segment with the highest mark.
(ii) the gaps between segments in the sequence of skeleton branches are filled. 
Starting from segments connected to extrema, all segments are visited iteratively: 
for the running segment, a mark is now assigned to all other unprocessed segments, 
based upon their relative distance, their relative angle, and relative orientation.
Note that the corresponding cost functions are non linear: for instance segments 
with too large  a relative angle are given an exponentially negative mark.
(iii)  finally, all segments which have not been considered during step (ii) are dropped.
 The resulting skeleton is
shown on figure \ref{fig:skl3D}.
A detailed accounting of all stages of the skeleton extraction, including 
the post treatment is given in \cite{TheseSousbie} (which gives the exact marking scheme described above), 
while the code is available upon request from the authors.\\

From a performance point of view, this method presents the
advantage of being both fast and robust.  The computational cost in
fact mainly scales as the number of pixels in grid $N_g^3$; the cost
of computing the iso surfaces intersections is neglectible given
the possibility of computing only the intersections of faces
belonging to the same pixel. It is moreover memory-efficient and can
be trivially parallelized: the computation can be done on sub grid
regions and then merged. On a modern computer, the memory requirement
corresponds to the requirement to store one sub-grid and its three isosurface,
which can be arbitrarily small, and the computational time for a
$128^3$ pixels grid is of the order of a few seconds on a modern desktop
computer while only a few tens of minutes is necessary for a grid of $1024^3$ pixels.

\begin{figure*}
\centering \subfigure[The three fields iso-surfaces the  intersection of which 
constitutes the
critical lines ]{\includegraphics[width=8cm,height=6.5cm]{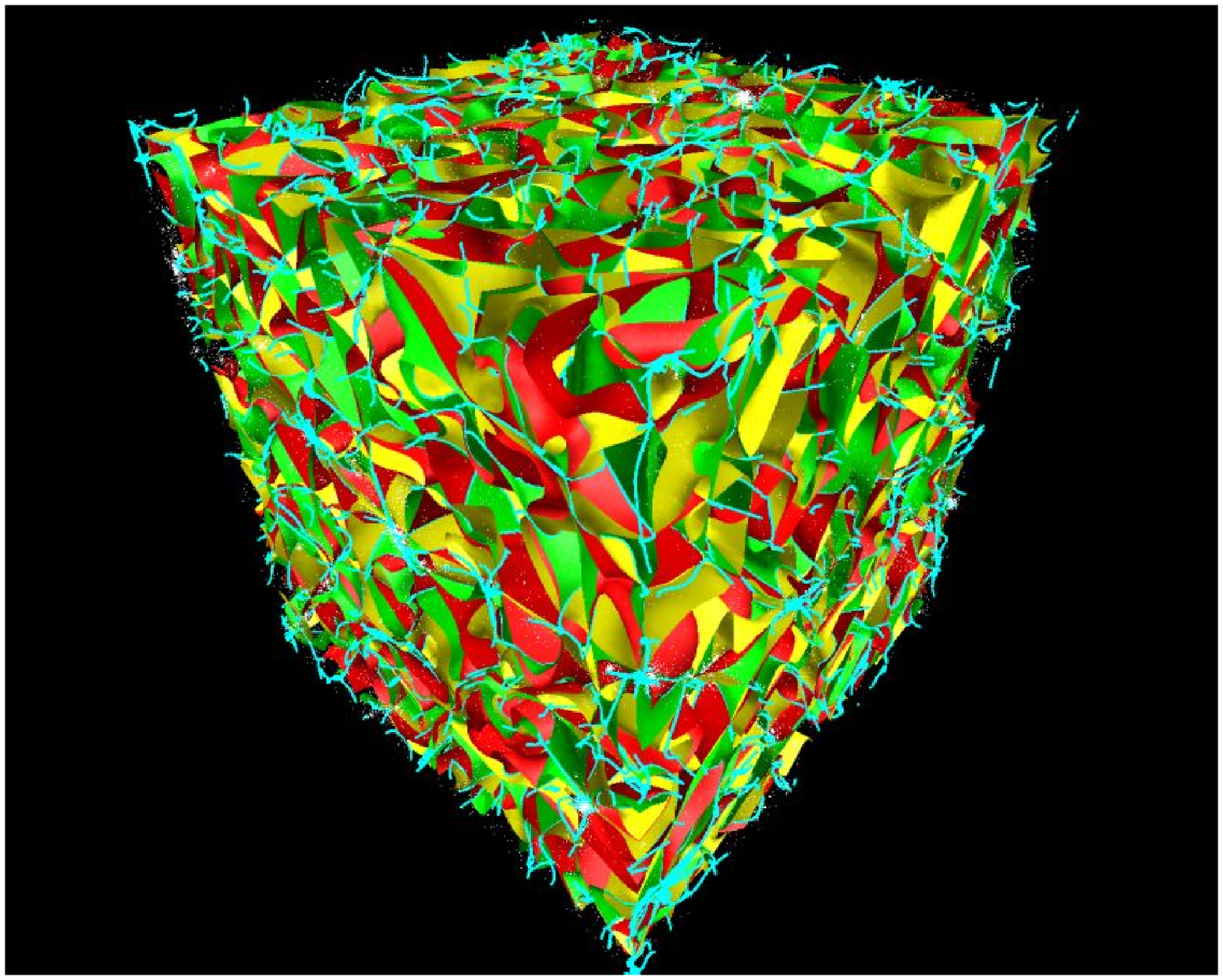}\label{fig:skl3D_evola}}
\hfill\subfigure[The resulting critical lines made  of all the intersection of
any two of the three iso-surfaces shown in
\ref{fig:skl3D_evola}.]{\includegraphics[width=8cm,height=6.5cm]{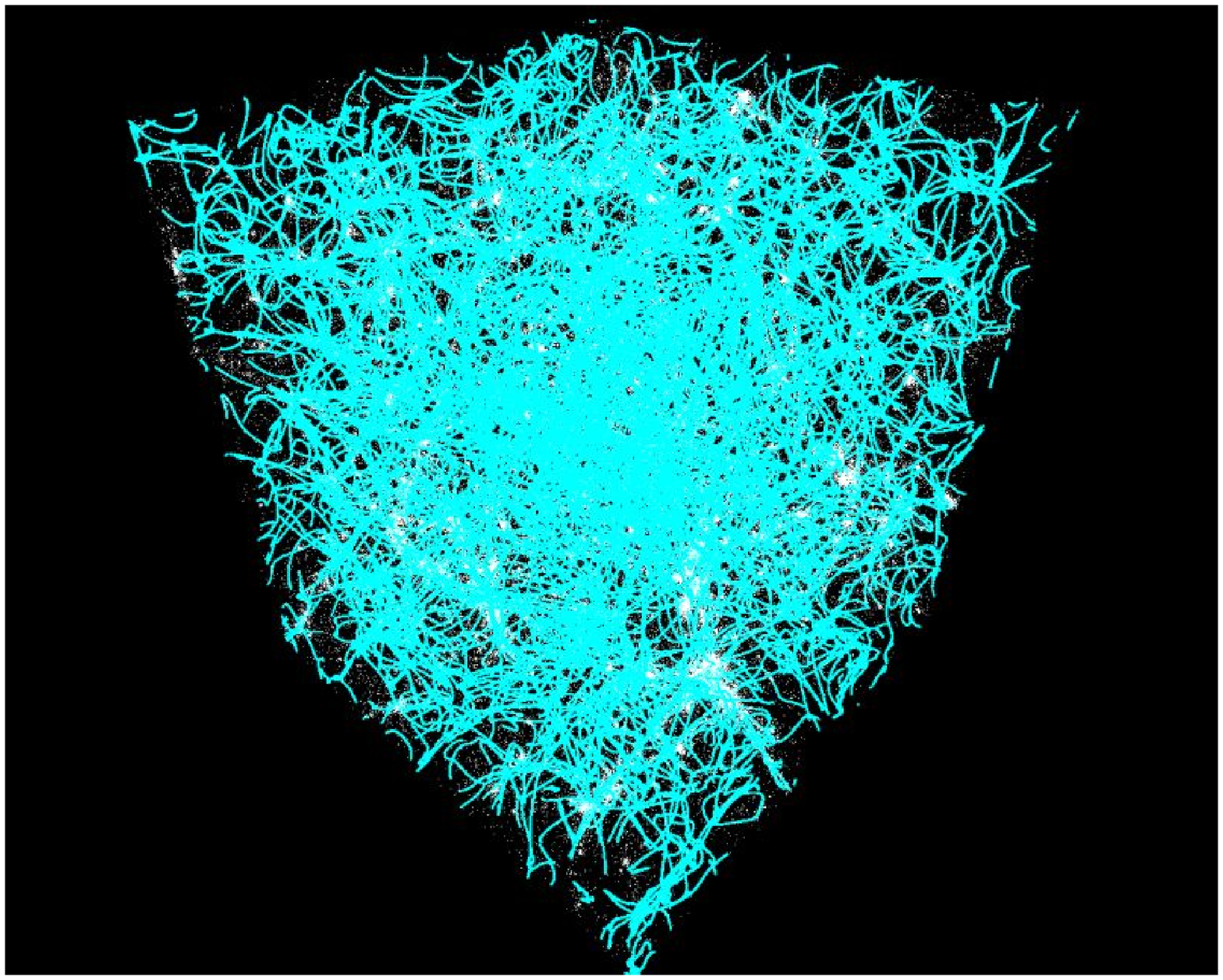}\label{fig:skl3D_evolb}}\\
\subfigure[The  local critical lines obtained by selecting only the two least degenerate fields
depending on the value of the gradient.]{\includegraphics[width=8cm,height=6.5cm]{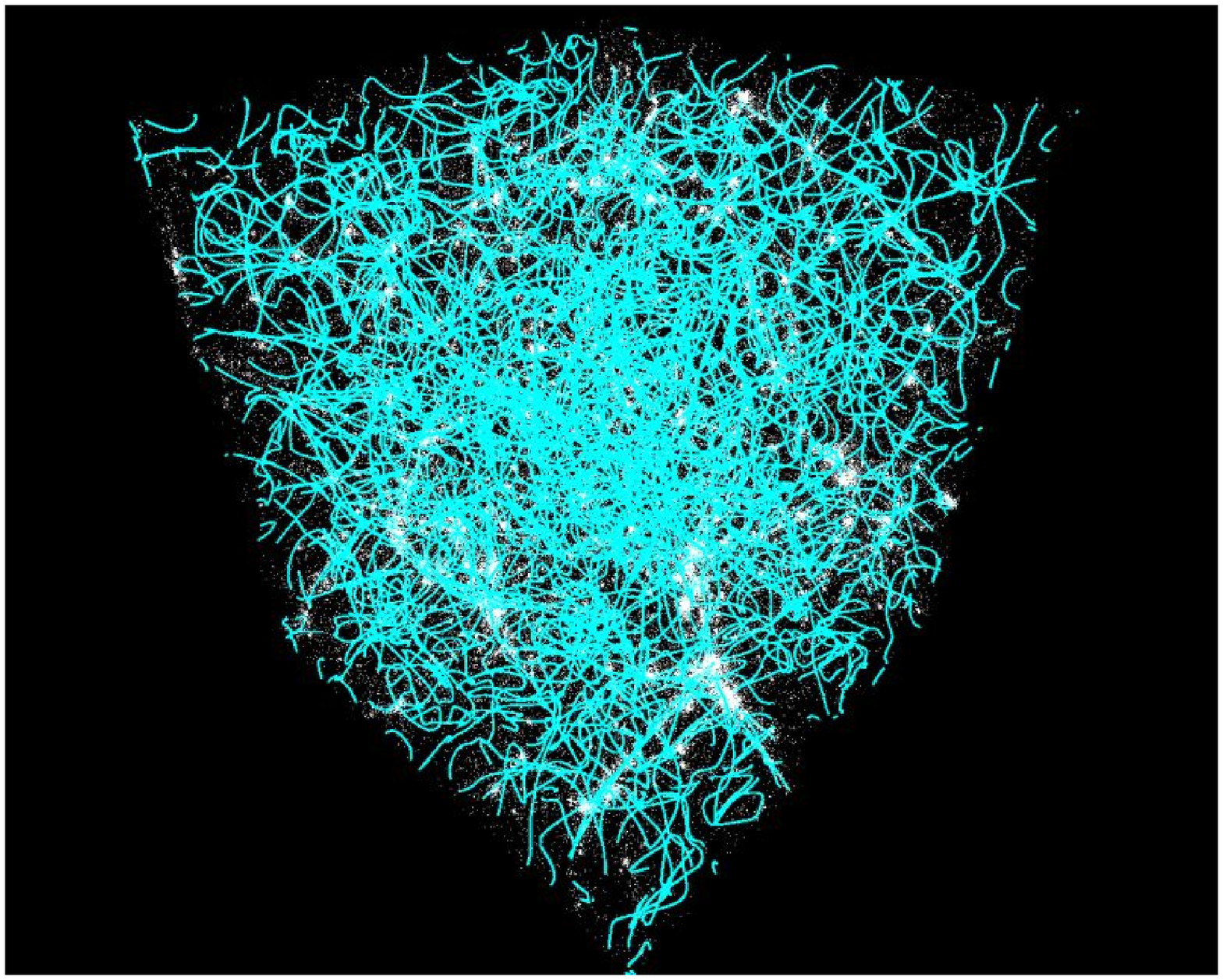}\label{fig:skl3D_evolc}}
\hfill\subfigure[The skeleton obtained after enforcing condition \ref{eq:eigselect3D} on the local critical lines: $\lambda_1>0$ and
$\lambda_2,\lambda_3<0$.]{\includegraphics[width=8cm,height=6.5cm]{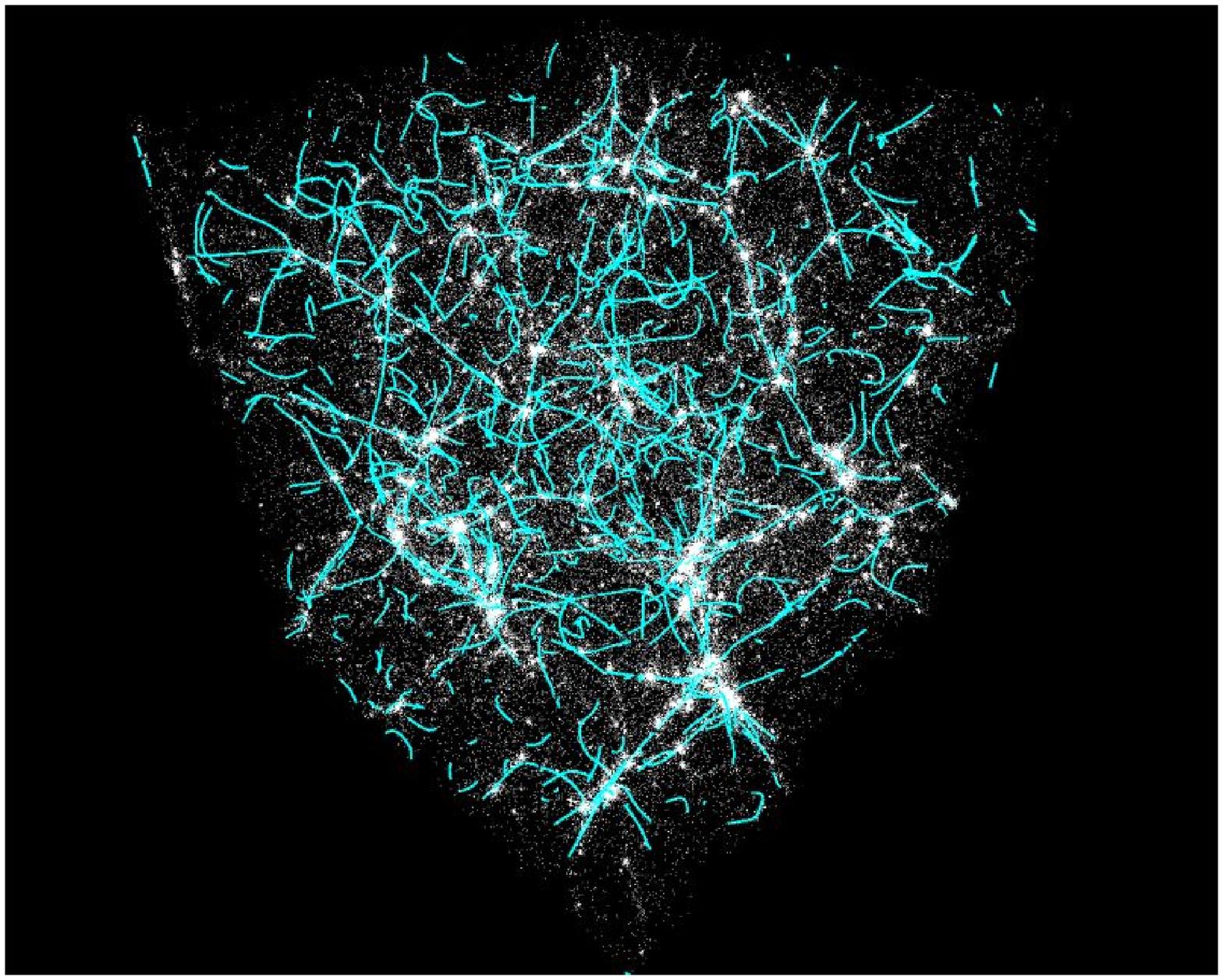}\label{fig:skl3D_evold}}
\caption{ Illustration of the process of the skeleton computation. White points
  are dark matter particles extracted from a standard $\Lambda$CDM simulation
  run using Gadget-2. The skeleton is defined as the intersection of two
  (among three) iso-surfaces (fig. \ref{fig:skl3D_evola}). Defining the
  curvature as $\lambda_i$ with ${\cal{H}}\nabla\rho=\lambda_i\nabla\rho$ and
  $\forall j>i$, $\lambda_j<\lambda_i$ ($\rho$ being the density and
  ${\cal{H}}$ its Hessian), it is possible to select only some parts of the
  skeleton depending on the value of  $\lambda_i$ and  retrieve only the
  filaments (fig. \ref{fig:skl3D_evold}). Using a simple post treatment, it is
  then possible to remove insignificant pieces and obtain the precise locus of
  the filaments (fig. \ref{fig:skl3D}).\label{fig:skl3D_evol}}
\end{figure*}

\section{Smoothing length and resolution}
\label{chap:slen}

\begin{figure}
  \centering \includegraphics[angle=0,width=7cm]{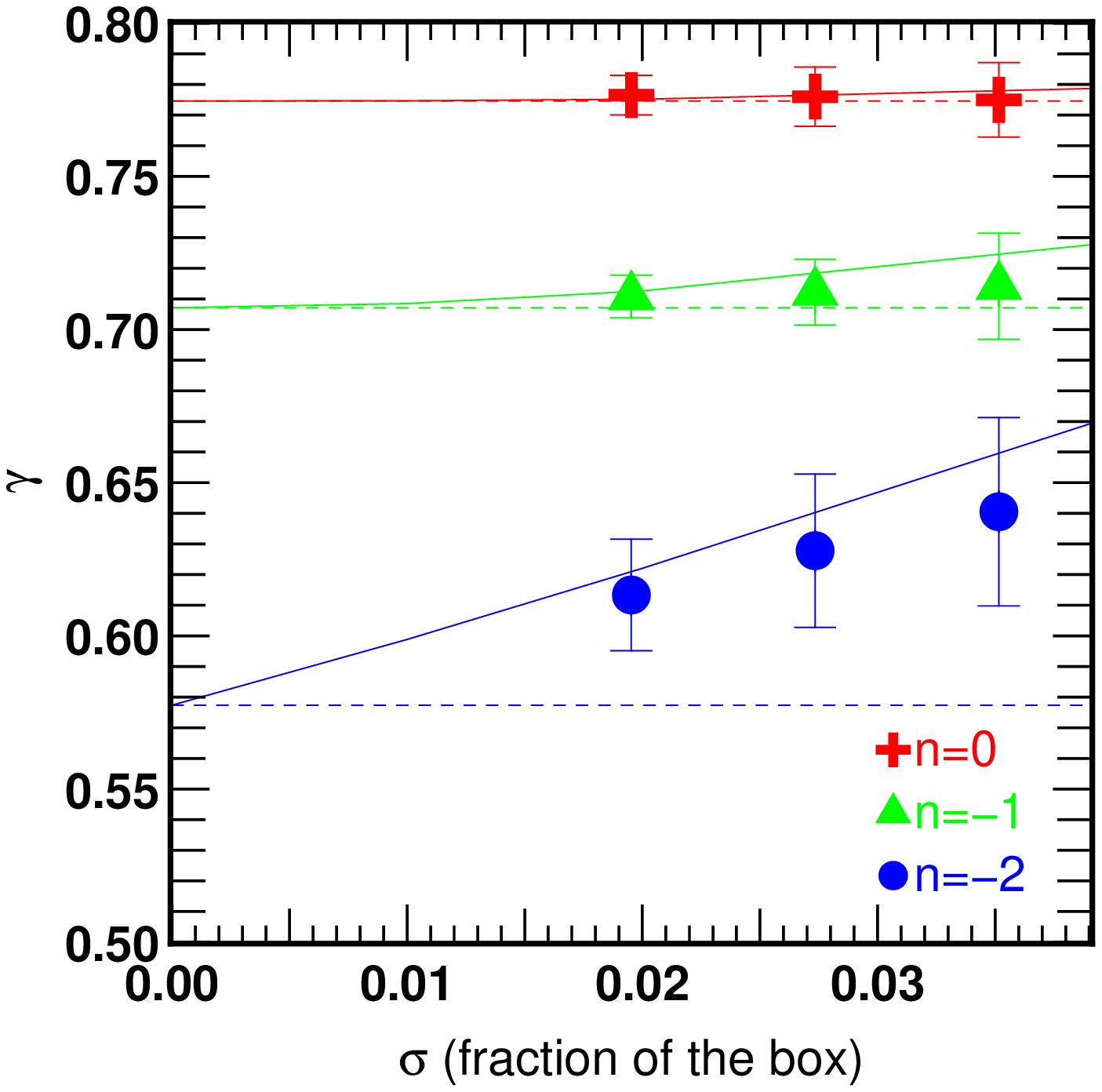}
   \centering \includegraphics[angle=0,width=7cm]{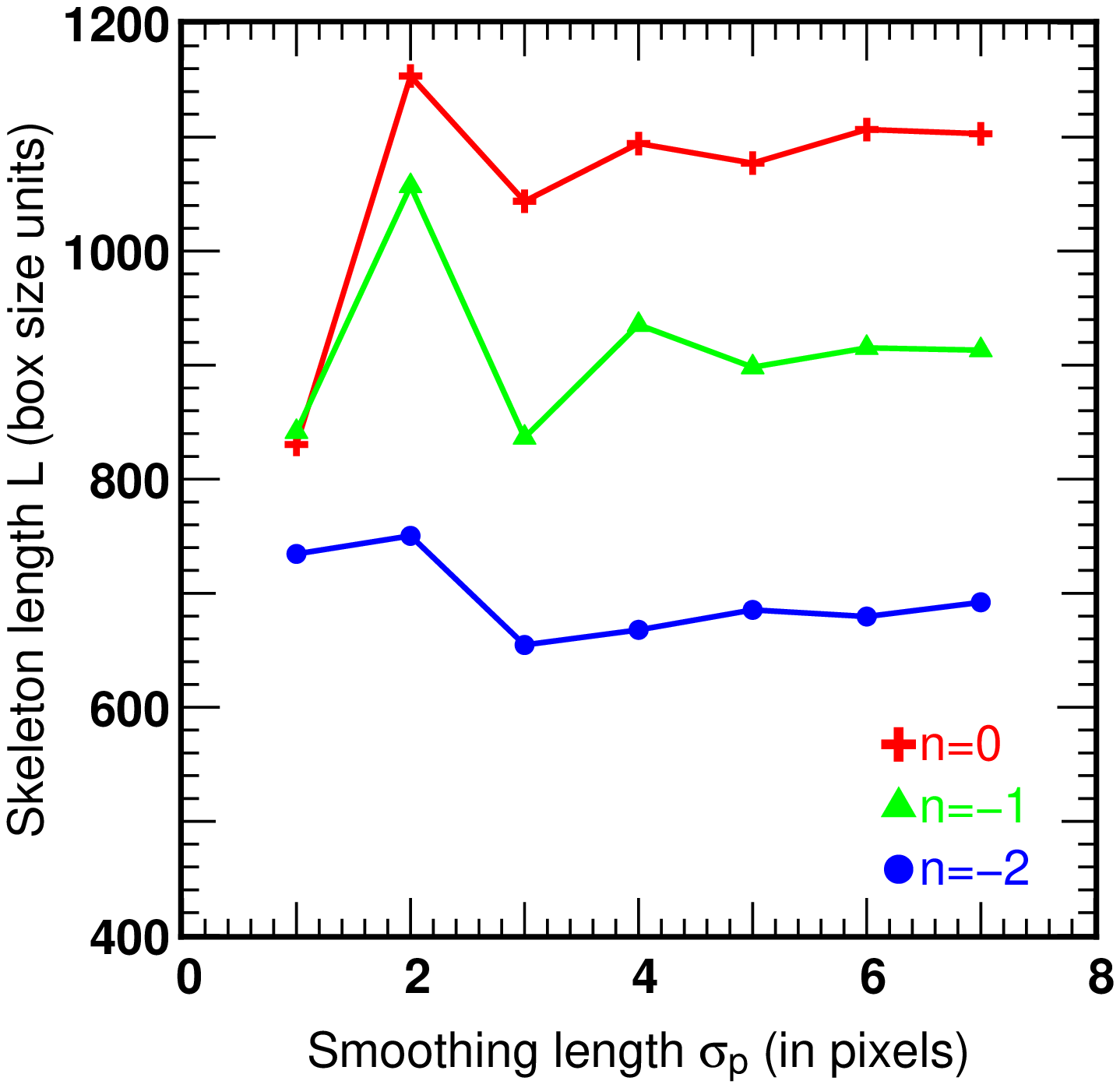}
 \caption{ {\sl Top: } evolution of the measured spectral parameter $\gamma$ (see Eq. \ref{eq:gammadef}) for
  $25$ realisations of Gaussian random fields with spectral index
  $n=0$ (red crosses), $n=-1$ (green triangles) and $n=-1$ (blue
  discs) as a function of the smoothing scale $\sigma$ expressed in
  box size units. The three continuous lines represent the expected
  theoretical values, measured by integrating the power spectrum truncated to grid limit frequencies.
  The dotted lines are the theoretical expectations (equation~(\ref{eq:thegamm})) without accounting for  finite volume effects.
   For higher values of $\sigma$, the finite box
  size effects have more influence and the measured value of $\gamma$
  tends to differ from the correct one, thus limiting the maximal
  smoothing scale
  {\sl Bottom}: evolution of the measured length of the total skeleton in
  box size units as a function of the smoothing length in pixels
  $\sigma_p$, for different values of the spectral index $n$ and while
  keeping the smoothing scale to a constant fraction of the box size
  $\sigma \approx 0.031$ . The measurements are obtained by resampling
  one initial realisation  of a Gaussian random field (genrerated over a $256^3$ pixels
  grid) on smaller resolution grids and smoothing the resulting fields
  over the appropriate number of pixels. The measured length of the
  total skeleton appears to become stable for values of $\sigma_p$
  above a limit of  $4$ to $5$ pixels at least, which corresponds to $\sigma>0.19$ 
  for a field sampled on a $256^3$ pixels grid.\label{fig:smoothtest}
  \label{fig:gamma}}
\end{figure}

One aspect of the numerical implementation that deserves special attention 
is the issue of smoothing.
In the main text, we consider the total skeleton
of Gaussian random fields, focusing mainly on two of its properties:
its length $L$ and differential length $dL/d\eta$. The algorithm
presented in this paper deals with the numerical computation of the
skeleton of a discretized realization of a given field. It is thus
important in the first instance to be able to deal with the influence of
this discretization on the measured skeleton properties (see e.g. Colombi \& al. 2000).\\

The statistical properties of a scale-free Gaussian random field can
be described using only two numbers: its spectral index $n$ and the
amplitude $A$ of its power spectrum $P\left(k\right)=Ak^n$, where $k$ is
the wave number. The skeleton formalism is totally independent of
the  amplitude of the field, so only the value of $n$ is of interest to
us. Consider a  realization of a 3D scale-free Gaussian random
field with spectral index $n$ on a $N_g^3$ pixel grid. In order to
ensure sufficient differentiability, this field is convolved  to a
Gaussian kernel whose scale $\sigma$ is expressed per unit box
size. The value of $\sigma$ limits the size of the  smallest scale
that can be considered, while the finite size of the grid imposes an
upper limit. Figure \ref{fig:gamma} presents the measured value of
the spectral parameter $\gamma^2=(n+3)/(n+5)$ as a  function of
$\sigma$, for $25$ realizations of Gaussian random fields with spectral
index $n\in\lbrace0,-1,-2\rbrace$,  together with the theoretical
value, measured by integrating the power spectrum truncated to grid limit frequencies. As expected, a departure from theory is observed for
higher values of $\sigma$,  especially for fields with lower spectral
index where most of the power is concentrated on small values of $k$ (i.e. on
large scales).  This sets an upper limit on the value of the smoothing
scale and so we will only be considering fields smoothed on scales
$\sigma\le 0.035$.\\

The other constraint on the value of $\sigma$ arises from the fact
that the skeleton computation algorithm requires a field that is
continuously differentiable  two times in the finite difference scheme
sense. This means that the smoothing length should be large enough for the computational errors on field derivatives to be
neglectible. These considerations imply  a lower limit on the smoothing
length value expressed in number of pixels $\sigma_p = \sigma N_g$. In
order to estimate this limit, we again generated Gaussian random fields with
different spectral indices over a $256^3$ ($N_g=256$) pixels grid and
downsampled them on grids with eight different values of $N_g$ ranging
from $N_g=64$ up to  $N_g=224$. Figure \ref{fig:smoothtest} presents
the  evolution of the measured skeleton length for these realisations,
each of them being computed for a smoothing scale corresponding to a
constant fraction of the box size  $\sigma \approx 0.031$ but to different values of $\sigma_p$ ranging from $\sigma_p=1$
up to $\sigma_p=7$. One would clearly expect the length of the
skeleton to depend only on the value  of $\sigma$ as long as the
numerical approximations are neglectible, which seems to be the case
only for values of  $\sigma_p$ at least of order $5$ pixels. For a
given sampling $N_g$, this limits the possible smoothing scale to
$\sigma > 5/N_g$. As was noted previously, this exact value depends 
on the considered spectral index, so we chose to consider the worst case, $n=1$, where 
the fluctuations of the field do not dampen on small scales thus making the field naturally not 
smooth on any scale.\\    
In this paper, all fields considered are sampled over $N_g=256$
cubic grids, so in order to respect the constraints described above,
 the fields  are smoothed on scales in the range
$0.02<\sigma<0.035$.

\newpage
\onecolumn

\section{The theoretical differential length of the skeleton}

\subsection{The length of the skeleton}
To estimate the length, ${\cal L}(\rho_{\rm th})$,
of the local skeleton per unit volume\footnote{the distinction is made here between the theoretical expectation, $\cal L(\rho_{\rm th})$, in this section and the 
estimator, $L$ in the main text.} consider the vicinity of the points through which  the local critical line passes, $S_i=0,S_j=0$. (Since the sets of conditions
$({\cal S}_i,{\cal S}_j)=(0,0)$, $i \neq j$ is degenerate, 
without loss of  generality one can assume  a particular choice  for $i$ and $j$).
Define the set  of points, $\cal E$, in the excursion $\rho > \rho_{\rm th}$ near the critical line solutions  that satisfy $-\Delta {\cal S}_i/2 \leq S_i \leq \Delta {\cal S}_i/2$
and $-\Delta {\cal S}_j/2 \leq S_j \leq \Delta {\cal S}_j/2$ where $\Delta S_i$ and $\Delta S_j$ are sufficiently small so that the linear expansion
$\Delta S_i \approx \nabla S_i \cdot {\mathbf dr} $,
 $\Delta S_j \approx \nabla S_j \cdot {\mathbf dr} $ holds\footnote{In such small neighborhood of a critical line there are no other critical lines. Note, that the linear expansion
 of $S_i$  breaks near the extrema of the field, where $\nabla S_i = 0$, which allows several critical lines to intersect at such points. However, extremal points are of measure zero as far as
 the computation of the length of the skeleton is concerned.} .
 The fraction of the total volume  the set $E$ occupies (the filling factor) is
 \begin{eqnarray}
{\cal V}(\rho_{\rm th},\Delta {\cal S}_i, \Delta {\cal S}_j) =
\int_{\rho > \rho_{\rm th}} d\rho \int_{-\Delta {\cal S}_i/2}^{\Delta {{\cal S}_i/2} }
d{\cal S}_i \int_{-\Delta {{\cal S}_j/2}}^{\Delta {{\cal S}_j/2}} 
d{\cal S}_j 
\int d^3(\nabla {\cal S}_i)\ d^3(\nabla {\cal S}_j) 
{\cal P}(\rho,{\cal S}_i,{\cal S}_j,\nabla {\cal S}_i,\nabla {\cal S}_j),
\label{eq:probdef}
\end{eqnarray}
where ${\cal P}(\rho,{\cal S}_i,{\cal S}_j,\nabla {\cal S}_i,\nabla {\cal S}_j)$ 
is the joint probability
distribution function of the quantities 
$(\rho,{\cal S}_i,{\cal S}_j,\nabla {\cal S}_i,\nabla {\cal S}_j)$.  Here the  seemingly redundant distribution of  the gradients $\nabla S_{i}$ and $\nabla S_{j}$ was introduced
to have the expression for the fraction of the total volume occupied
 by a differential subset of $\cal E$ that has specific values of the gradients $\nabla S_{i}, \nabla S_{j}$ (within $ d^3(\nabla {\cal S}_i)$
and $\ d^3(\nabla {\cal S}_j )$)
\begin{eqnarray}
d{\cal V}(\rho_{\rm th},\Delta {\cal S}_i, \Delta {\cal S}_j,\nabla {\cal S}_i ,\nabla {\cal S}_j) = d^3(\nabla {\cal S}_i)\ d^3(\nabla {\cal S}_j)
\int_{\rho > \rho_{\rm th}} d\rho \int_{-\Delta {\cal S}_i/2}^{\Delta {{\cal S}_i/2} }
d{\cal S}_i \int_{-\Delta {{\cal S}_j/2}}^{\Delta {{\cal S}_j/2}} 
d{\cal S}_j 
{\cal P}(\rho,{\cal S}_i,{\cal S}_j,\nabla {\cal S}_i,\nabla {\cal S}_j)  \,.
\label{eq:probdef}
\end{eqnarray}
Since the  area, $\Sigma$, of a section locally orthogonal
to the such subset, is simply (modulo some trigonometry) given by 
\[
\Sigma(\Delta {\cal S}_i,\Delta {\cal S}_j,\nabla {\cal S}_i ,\nabla {\cal S}_j)=\Delta {\cal S}_i \Delta {\cal S}_j /|\nabla {\cal S}_i \times \nabla {\cal S}_j|,
\] 
dividing $d {\cal V}$ by $\Sigma$,  integrating over all possible gradients $\nabla S_{i},\nabla S_{j}$ and then taking the
limit $(\Delta {\cal S}_i,\Delta {\cal S}_j) \rightarrow (0,0)$  yields for the skeleton length per unit volume:  
\begin{eqnarray}
{\cal L}(\rho_{\rm th})&=&\!\!\!\!\!\!\!\!\!\!\!\!\!\!
\lim_{(\Delta {\cal S}_i,\Delta {\cal S}_j) \to (0,0)} \int  \frac{d {\cal V}(\rho_{\rm th},\Delta {\cal S}_i, \Delta {\cal S}_j,\nabla {\cal S}_i ,\nabla {\cal S}_j)}{\Sigma(\Delta {\cal S}_i,\Delta {\cal S}_j,\nabla {\cal S}_i ,\nabla {\cal S}_j)} 
 \nonumber \\ &=&
  \int_{\rho > \rho_{\rm th}}\!\!\!\!\!\!\!  d\rho\  \int d^3(\nabla {\cal S}_i)\
d^3(\nabla {\cal S}_j)\ |\nabla {\cal S}_i \times \nabla {\cal S}_j|  
{\cal P}(\rho,{\cal S}_i=0,{\cal S}_j=0,\nabla {\cal S}_i,\nabla {\cal S}_j). \label{eq:defL0}
\end{eqnarray}
This generalizes the calculation of NCD to 3 dimensions: 
the length of the local skeleton is defined by the knowledge of the density field and its partial derivatives
up to third order,  as expected. 
In order to understand the scaling involved in the computation of $\cal L$,
let us  rewrite this equation in terms of dimensionless quantities:
\begin{eqnarray}
\sigma_0 x \equiv \rho,\quad \sigma_1 x_{i}\equiv \frac{\partial \rho}{\partial r_i},
\quad
\ \sigma_2 x_{ij}\equiv \frac{\partial^2 \rho}{\partial r_i \partial r_j},\quad 
\sigma_3 x_{ijk}\equiv\frac{\partial^3 \rho}{\partial r_i \partial r_j \partial r_k}, 
\quad
\sigma_2 \sigma_1^2 s_i \equiv {\cal S}_i,\quad \sigma_3 \sigma_1^2 \nabla s_i \equiv \nabla {\cal S}_i,
\label{eq:scalefr} \end{eqnarray}
with, following \cite{BBKS},
\begin{equation}
\sigma_n^2 \equiv \int \frac{k^2 dk}{2 \pi^2} P(k) k^{2n},
\end{equation}
where $P(k)$ is the power-spectrum of $\rho$. 
Equation~(\ref{eq:defS2})  and its gradient can be written more conveniently, using the totally antisymmetric tensor, $\epsilon^{ijk} $, as 
\begin{equation}
s_i= \sum_{jkl} \epsilon^{ijk} x_{jl} x_{l} x_{k}\,, \quad {\rm  and} \quad \nabla_{m} s_i\equiv \nabla {\hat s}_i(x_{k},x_{kl},x_{klm})= 
\sum_{jkl} \epsilon^{ijk} \left( x_{jlm} x_{l} x_{k}+{\tilde \gamma}\left[ x_{jl} x_{lm} x_{k}+ x_{jl} x_{km} x_{l}
\right]\right)\,.
\label{eq:defsi}
\end{equation} 
  Indeed, expressions for $s_{i}$ and $\nabla s_{i}$ involve up to third derivatives of the field $x$.
All the quantities defined by Eq.~(\ref{eq:scalefr}) are
dimensionless and their variance do not depend on spectral parameters (they are pure numbers)
except $\nabla s_i$. 
Keeping that in mind, with these new notations, equation~(\ref{eq:defL0}) becomes 
\begin{eqnarray}
{\cal L}(\rho_{\rm th})=\left( \frac{\sigma_3}{\sigma_2}\right)^2 \int_{x > x_{\rm th}} dx\int \ d^3(\nabla s_i)\
d^3 (\nabla s_j)\ |\nabla s_i \times \nabla s_j| 
{\cal P}(x,s_i=0,s_j=0,\nabla s_i,\nabla s_j). \label{eq:myeq}
\end{eqnarray}
Equation~(\ref{eq:myeq}) is the formal expression for the length per unit volume of the total set of critical lines.

Let us express the joint distribution function ${\cal P}(\eta,s_i,s_j,\nabla s_i,\nabla s_j)$ in terms of the joint distribution function of the
underlying field and its derivatives $P(x,x_{k},x_{km},\ldots)$.
Introducing the 20 components vector composed of 
$x$ and its successive partial derivatives
up to third order (see equation~\ref{eq:scalefr}), or in dimensionless units, ${\bf X}\equiv (x,x_k,x_{kl},x_{klm})$
(symmetries in the derivative tensors of second and third order are assumed to be exploited
to reduce the effective number of variables), we can find $\cal P$ as a marginalization over the distribution of the field values:
\begin{eqnarray}
{\cal P}(\eta,s_i,s_j,\nabla s_i,\nabla s_j)= \int^{\eta} d x  d^{3}x_{k} d^{6} x_{kl}  d^{10}x_{klm}   {P}(x,x_{k},x_{kl},x_{klm} )\delta_{D}(x-\eta) \delta_{D}( {\hat s}_{i}(x_{k},x_{kl})-s_{i})\delta_{D}( {\hat s}_{j}(x_{k},x_{kl})-s_{j})\times \nonumber \\
\delta_{D}( \nabla {\hat s}_{i}(x_{k},x_{kl},x_{klm})- \nabla s_{i})\delta_{D}(  \nabla {\hat s}_{j}(x_{k},x_{kl},x_{klm})-  \nabla s_{j})\,,
\label{eq:P2P}
\end{eqnarray}
which yields the appropriate 9D probability distribution.
Putting equation~(\ref{eq:P2P}) into equation~(\ref{eq:myeq}) and differentiating with respect to $\eta=\rho/\sigma_{0}=x$, and accounting for the two Delta functions 
in $\nabla s_{i}$ and $\nabla s_{j}$ yields, using equation~(\ref{eq:gammadef}) to rewrite ${\sigma_3}/{\sigma_2}$ in terms of $\tilde R$: 
\begin{eqnarray}
\frac{\partial{\cal L}}{\partial \eta}=\left( \frac{1}{\tilde R}\right)^2 \int  d^{3}x_{k} d^{6} x_{kl}  d^{10} x_{klm} \ |\nabla {\hat s}_i \times \nabla {\hat s}_j|
{P}(\eta,x_{k},x_{kl},x_{klm}) \delta_{D}( {\hat s}_{i}(x_{k},x_{kl}))\delta_{D}( {\hat s}_{j}(x_{k},x_{kl})) \,.
 \label{eq:myeq2}
\end{eqnarray}
Equation~(\ref{eq:myeq2}) is the formal expression for the differential length per unit volume of the total set of critical lines, described in the main text.
Given its dimensionality, it remains a daunting task to compute the 19D  integral in  equation~ (\ref{eq:myeq2}). 
Note that $\nabla {\hat s}_i$ and $\nabla {\hat s}_j$ are now functions of $(x_{k},x_{kl},x_{klm})$  and 
 ${\hat s}_{i}$ and  ${\hat s}_{j}$ are function of $(x_{k},x_{kl})$  given by equation~(\ref{eq:defsi}).
In equation~(\ref{eq:myeq2}), the two Delta functions couple the differents $x_{k},x_{kl}$ while accounting for the fact that the integral should be restricted 
to the intersection of the two iso surfaces, i.e. along the critical lines. The modulus in $|\nabla {\hat s}_i \times \nabla {\hat s}_j|$ reflects the fact 
that the summation of skeleton segments is not algebraic, which complicates also the reduction of equation~(\ref{eq:myeq2}). For the  set of local critical lines, there are no restriction to the region of integration. If one is interested in the local skeleton, the integration should be restricted to 
regions where the  condition  given by equation (\ref{eq:eigen}) holds.

The total length of the critical lines is
\begin{equation}
{\cal L}_{tot} = \int_{-\infty}^{\infty} d \eta \frac{\partial{\cal L}}{\partial \eta}=\left( \frac{1}{\tilde R}\right)^2 \int  d^{3}x_{k} d^{6} x_{kl}  d^{10} x_{klm} \ |\nabla {\hat s}_i \times \nabla {\hat s}_j|
{P}(x_{k},x_{kl},x_{klm}) \delta_{D}( {\hat s}_{i}(x_{k},x_{kl}))\delta_{D}( {\hat s}_{j}(x_{k},x_{kl})) \,.
\end{equation}

\subsection{${\cal L}(\rho_{\rm th})$ for Gaussian random field}

Since a Gaussian field does not correlate with its  derivatives of
odd orders (this is easy to understand using symmetries in Fourier space),
the joint distribution function $P(x,x_{k},x_{kl},x_{klm})$ can be written as 
\begin{equation}
{P}(x,x_{k},x_{kl},x_{klm})={ P}_{0}(x,x_{kl}) {P}_{1}(x_{k},x_{klm}).
\end{equation}
In $P_0$, the only dependence on the power spectrum of the underlying field  is in the parameter $\gamma$ (c.f. equation~(\ref{eq:gammadef}))
that describes the correlation between the field and its second derivatives.
Similarly  ${P_{1}}(x_{i},x_{ijk})$ only involves $\tilde \gamma$ which describes the correlation between the gradient of the field and its third derivatives.
Therefore ${\partial{\cal L}}/{\partial \eta}$ depends only on $\eta$, $\tilde R$ $\gamma$  and $\tilde \gamma$, as argued in the main text.
 Note that by symmetry, ${\partial{\cal L}}/{\partial \eta}$ should then be an even function of $\eta$ for the total 
 set of critical lines.
The total length of the skeleton, ${\cal L}_{\rm tot}$, which follows from marginalization of the  equation~(\ref{eq:myeq2}) over $\eta$ 
may depend only on  $\tilde \gamma$ and $\tilde R$ since the integration of $P_{0}(\eta,x_{kl})$ over $\eta$ cancels the dependancy over $\gamma$.

\subsubsection{The ``stiff'' filament approximation}

The $1/\tilde R^{2}$ scaling in equation~(\ref{eq:myeq2})
 reflects the basic fact that by definition the local skeleton is almost straight within a volume that has one inflection point $\sim \tilde R^{3}$. A straight segment through such volume has
length $\sim \tilde R$, thus the expected length per unit volume is $\sim 1 /\tilde R^{2}$. The dependence on the spectral index  is then $1/\tilde R^{2} \propto (n+7) /\sigma^{2}$,
recalling that $\sigma$ is the smoothing length in units of the total boxsize.  
Is this the behaviour with $n$ that one should expect in simulations ?
\\
Let us write formally 
\begin{equation}
\nabla s_{i}\times \nabla s_{j} = {\mathbf A}(x_{k},x_{kl},x_{klm}) +{\tilde \gamma} {\mathbf B} (x_{k},x_{kl},x_{klm})  +{\tilde \gamma}^{2} {\mathbf C} (x_{k,}x_{kl}) \,.\label{eq:simple}
\end{equation}
Suppose the last term dominates statistically\footnote{or equivalently assume that the amplitude of  derivative of the Hessian is negligible relative to the amplitude of the Hessian}. 
 Then, since $\tilde \gamma/\tilde R = 1/R_{*}$, and given that $\mathbf{C} (x_{k,}x_{lm}) $ does not depend on the third derivative of the field 
(which can then be integrated out), equation~(\ref{eq:simple}) becomes
\begin{equation}
\frac{\partial{\cal L}}{\partial \eta} \approx \left( \frac{1}{R_{*}}\right)^2 \int d^{3}x_{k} d^{6} x_{kl}\ | {\mathbf C}(x_{k,}x_{lm}) |
\ {P}_{0}(\eta,x_{kl}) \  P_{1}(x_{k}) \delta_{D}( {\hat s}_{i}(x_{k},x_{kl}))\delta_{D}( {\hat s}_{j}(x_{k},x_{kl})) \,. \label{eq:dldetastiff}
\end{equation}
It is easy to foresee when this regime is valid.  The same argument as before implies that the  $1/R_{*}^{2}$ scaling arises when the skeleton is almost straight within a volume that contains one extremum, $R_{*}^{3}$. This is supported by the fact that the integral term does not depend on the third derivatives, thus inflection points play no role, and any dependence on $\tilde\gamma$ drops out. This picture corresponds to a  skeleton connecting extrema with relatively straight segments. The scaling  is then $1/R_{*}^{2}\propto (n+5)/\sigma^{2}$.

For the total length of the critical lines, integration over $\eta$ gives
\begin{equation}
{\cal L}_{tot} \approx \left( \frac{1}{R_{*}}\right)^2 \int d^{3}x_{k} d^{6} x_{kl}\ | {\mathbf C}(x_{k,}x_{lm}) |
\ {P}_{0}(x_{kl}) P_{1}(x_{k}) \delta_{D}( {\hat s}_{i}(x_{k},x_{kl}))\delta_{D}( {\hat s}_{j}(x_{k},x_{kl})) \propto (n+5) \sigma^{-2}\,,
\end{equation}
since the integral is just a pure number.  This is very close to the scaling with $n$ that was found in the numerical fit, equation~(\ref{eq:defLL}).
The differential length in the stiff regime is then only the function of $\gamma$ times ${\cal L}_{tot}$.
The upshot of this paragraph is to demonstrate the theoretical consistency between the scaling in $ (n+5) \sigma^{-2}$ of ${\cal L}_{tot} $
and the fact that ${\partial{\cal L}}/{\partial \eta} $ does not depend on $\tilde \gamma$ for scale free Gaussian random fields.

\subsubsection{Joint distribution of the field and its derivatives for a Gaussian random field}

The full expression
 ${P}_{0}(x,x_{kl})$ is given in \cite{BBKS}. 
 Introducing variables
\begin{equation}
u  \equiv - \Delta x = -(x_{11}+x_{22}+x_{33})\,,\quad
v  \equiv\frac{1}{2}  (x_{33}-x_{11})\,,  \quad
w  \equiv   \sqrt\frac{1}{12}(2 x_{22}-x_{11} - x_{33})\,,  \label{eq:defu}
\end{equation}
in place of diagonal elements of the Hessian $(x_{11},x_{22},x_{33})$ one finds that
$u,v,w,x_{12} ,x_{13},x_{23}$ are uncorrelated. Importantly, the field, $x$ is only correlated with $u=\Delta x$ and
 \begin{equation}
\langle x u \rangle= \gamma, \quad \langle x v \rangle =0, \quad 
\langle x w \rangle=0, \quad \langle x x_{kl} \rangle=0, \ k\neq l,
\end{equation}
where $\gamma$ is the same quantity as in equation~(\ref{eq:gammadef}).
The full expression of ${P}_{0}(x,x_{kl})$ is then
\begin{eqnarray}
P_{0}(x,x_{kl}) dx d^6 x_{kl}=\frac{(15)^{5/2}}{(2\pi)^{7/2}({1-\gamma^2})^{1/2}} \exp\left[ 
-\frac{(x -\gamma u)^2}{2(1-\gamma^2)} -\frac{u^2}{2}\right] \exp\left[-\frac{15}{2}({v^{2}+w^{2}+x_{12}^{2}+x_{13}^{2}+x_{23}^{2}})\right]
dx du  dv dw dx_{12}  dx_{13}  dx_{23}, \nonumber
\end{eqnarray}
and is described by only one correlation parameter $\gamma$.

A similar procedure  can be performed for the joint probability of the first and third derivatives of the fields, ${P_{1}}(x_{i},x_{ijk})$
by defining the following nine parameters (see also Hanami  99):
\begin{equation}
u_i \equiv \nabla_{i} u,
\quad v_i \equiv \frac{1}{2} \epsilon^{ijk} \nabla_{i } \left( \nabla_{j} \nabla_{j}- \nabla_{k} \nabla_{k}\right)x \,,\,\,\, {\rm with}\,\,\, j<k\,,\quad {\rm and }
\quad w_i \equiv \sqrt\frac{5}{12}\nabla_{i}\left( \nabla_{i} \nabla_{i}  -\frac{3}{5}\Delta \right)x\,,
\end{equation}
and replacing the variables $(x_{i11},x_{i22},x_{i33})$ with
$(u_i,v_i,w_i)$. 
In that case, the only cross-correlations in the vector $(x_1,x_2,x_3,u_1,v_1,w_1,u_2,v_2,w_2,
u_3,v_3,w_3,x_{123})$ which 
do not vanish are between the same components of the gradient and the gradient of the Laplacian of the field:
\begin{eqnarray}
\langle x_i u_i \rangle & = & {\tilde \gamma}/3,\quad i=1,2,3,
\end{eqnarray}
where $\tilde \gamma$ is the same quantity as in equation~(\ref{eq:gammadef}).
This allows us to write:
\begin{eqnarray}
{P_{1}}(x_{i},x_{ijk}) d^3x_i\, d^{10} x_{ijk}=\frac{105^{7/2} 3^{3} d^{3}w_i \, d^{3}v_i\,dx_{123}}{(2\pi)^{13/2}(1-{\tilde \gamma}^2)^{3/2}} 
 \exp\left[-\frac{105}{2}\left(
 x_{123}^{2}+\sum_{i=1}^{3}(v_{i}^{2}+w_{i}^{2})\right)
 \right]
\prod_{i=1}^3 du_i dx_i 
\exp\left[-\frac{3(u_i -{\tilde \gamma} x_i)^2}{2(1-{\tilde \gamma}^2)} -\frac{3 x_i^2}{2}\right] 
. \nonumber
\end{eqnarray}
\subsubsection{The differential length for Gaussian fields }
What is the dependence of the skeleton differential length on the parameter $\gamma$ and the threshold $\eta$?  Let us look at the structure of the integrals involved with respect to
the variable $u$.  Importantly,  the arguments of the Delta functions $S_i \ne S_i(u)$ and $\nabla S_i \times \nabla S_j$, given by equation~(\ref{eq:simple}),  is  $ \sim \sqrt{Q_4(u)} $  where $Q_{4}(u)$ is a positive quartic in $u$. 
Puting the expressions for $P_{1}$ and $P_{0}$ into equation~(\ref{eq:myeq2})  imply that 
the $u$-component of the integral in ${\partial{\cal L}}/{\partial \eta}$  involves 
\begin{equation}
{\cal I}(\gamma,\eta)=\int \limits_{-\infty}^{\infty}
\frac{ \sqrt{Q_{4}(u)}  \exp\left(-{u^2}/{2} \right) }{(2\pi)^{1/2}({1-\gamma^2})^{1/2}} \exp\left[ 
-\frac{(\eta-\gamma u)^2}{2(1-\gamma^2)}\right] d u\,, \label{eq:core}
\end{equation}
where $Q_4(u)$, of course, also depends on $v,w,u_{k},v_{k},w_{k}, x_{123}, x_{kl}$ with $ k< l$ and possibly $\tilde \gamma$, but not on $\gamma$.  

In the trivial limit $\gamma \to 0$ the coupling between $u$ and the field value $\eta$ vanishes and the differential length is reduced to the PDF of $\eta$
\begin{equation}
d {\cal L}/d \eta \propto \exp\left[-\frac{\eta^2}{2}\right] =\frac{ {\cal L}_{tot}}{(2\pi)^{1/2}} \exp\left[-\frac{\eta^2}{2}\right]
\end{equation}

For nonvanishing $\gamma$,  following NCD, the differentiation  of equation~(\ref{eq:core}) shows that 
${\cal I}(\gamma,\eta)$ obeys the equation
\[
\gamma \frac{\partial {\cal I}}{\partial \gamma } =-\frac{\partial}{\partial \eta } \left[ \eta {\cal I}(\gamma,\eta)+ \frac{\partial {\cal I}}{\partial \eta } 
\right]\,, \label{eq:eqdiff}
\]
whose solution involve even Hermitte polynomials (retaining only the convergent solution at large $\eta$):
\begin{equation}
{\cal I}(\gamma,\eta)=\sum_{n=0}^{\infty} c_{ 2n} \gamma^{2n} H_{2n}( \eta/\sqrt{2} ) \exp\left[  -\frac{\eta^2}{2}\right] \,. \label{eq:core3}
\end{equation}
Note that $c_{0}$ is non null since asymptotically, the differential length should converge towards the PDF of $\eta$, within a multiplicative constant.
Thanks to the orthogonality condition on the Hermitte polynomial, 
$c_{2n}$ is given by
\begin{equation}
c_{2n}=\lim_{\gamma \to 1}\int d x H_{2 n }(x/\sqrt{2}) \exp(-x^{2}/2) {\cal I}(x,\gamma)=\int d x  H_{2 n }(x/\sqrt{2}) \exp(-x^{2})\sqrt{Q_{4}(x)}=c_{2n}(v,w,u_{i},u_{i},w_{i}, x_{123}, x_{ij})\,. \label{eq:defcn}
\end{equation}
The integration of equation~(\ref{eq:core3}) over $v,w,u_{k},u_{k},w_{k}, x_{123}, x_{kl}$  for $ k< l$ (while accounting 
for the rest of the integrant corresponding to $P_{0}$ and $P_{1}$ together with the two Delta functions) yields the functional form of $d {\cal L}/d \eta$, equation~(\ref{eq:fitdldn}), where 
$C_{2n}$ is a pure number in the stiff approximation, but may depend on $\tilde \gamma$ in general.
This section falls short of demonstrating why  only $C_{2}$ is non null, though the oscillatory behaviour of $H_{2n}$ in equation~(\ref{eq:defcn}) suggests that $C_{2n}$ 
should decrease with $n$.

Making use of the expression of $P_{0}$ and $P_{1}$ in the stiff regime ($x_{klm}=0$), equation~(\ref{eq:dldetastiff}) becomes
\[
\frac{\partial{\cal L}}{\partial \eta} \propto \int
\frac{ |Q_{2}(u)|}{({1-\gamma^2})^{1/2}} \exp\left[ 
-\frac{(u -\gamma \eta)^2}{2(1-\gamma^2)} -\frac{\eta^2}{2}\right] du \exp\left[-\frac{15}{2}({{\hat v}^{2}[x_{kl}]+{\hat w}^{2}[x_{kl}]+\sum_{k<l} x_{kl}^{2})-\frac{3}{2}\sum_{k} x_{k}^{2}}\right]
 \left|\frac{\partial ( v,w) }{\partial ({s}_{i}, {s}_{j}) } \right| 
\prod_{k<l}  dx_{kl}  \prod_{k} dx_{k},
\]
where 
${\hat v}^{2}[x_{jk},s_{i}=0,s_{j}=0]$ and ${\hat w}^{2}[x_{jk},s_{i}=0,s_{j}=0]$ are function 
of $x_{kl}$, $k<l$  and  $s_{i}$ and $s_{j}$ (evaluated at zero to account
for the two Delta functions).
The quadratic function $Q_{2}(u)$ corresponds to the square root of 
$Q_{4}(u)$ when $v$ and $w$ are reexpressed in terms of $s_{i}$ (which are in turn 
evaluated at zero).
Now writing formally again $Q_{2}(u) = A_{2} u^{2}+ A_{1} u +A_{0}$, if the region where $Q_{2}(u)$ is positive dominates the above integral,
the integration over $u$ yields
\[
\left[A_{0} +A_{2}+ A_{1} \gamma \eta + A_{2}\, \gamma^{2}\, H_{2}(\eta/\sqrt{2})\right]
\exp(-\eta^{2}/2) 
\]
which suggests that $C_{2n}=0 $ for $n>1$. Given that ${\partial{\cal L}}/{\partial \eta} $ is even in $\eta$, here  $A_{1} \gamma$ should eventually cancel out when integrated
over the other variables.


\begin{thebibliography}{}
\bibitem[Alcock \& Paczynski 1979]{Alcock} Alcock C., Paczynski B., 1979, 
Natur, 281, 358
\bibitem[Aubert \& Pichon(2007)]{2007MNRAS.374..877A} Aubert, D., \& 
Pichon, C.\ 2007, \mnras, 374, 877 
\bibitem[Aubert, Pichon \& Colombi (2004)]{2004MNRAS.352..376A} Aubert, D., 
Pichon, C.  \& Colombi, S. \ 2004, \mnras, 352, 376A 
\bibitem[Bardeen, et al., 1986]{BBKS}  Bardeen, J. M., Bond, J. R., Kaiser, N., Szalay, A. S., 1986, ApJ 304, 15 (BBKS)
\bibitem{} Barrow, J. D., Bhavsar, S. P., Sonoda, D. H., MNRAS 216, 17
\bibitem{} Bertschinger E., 1985, ApJS, 58, 1
\bibitem{} Bond, J. R., Kofman, L. A., Pogosyan, D, 1996, Nature, 380, 63.
\bibitem{} Bond, J. R., Myers, S. T., 1996a, ApJS 103, 1
\bibitem{} Bond, J. R., Myers, S. T., 1996b, ApJS 103, 41
\bibitem[Colless et al.(2003)]{2003astro.ph..6581C} Colless, M., et al.\ 
2003, ArXiv Astrophysics e-prints, arXiv:astro-ph/0306581 
\bibitem{} Colombi, S., Pogosyan, D., Souradeep, T., 2000, Phys. Rev. Lett.
  85, 5515
\bibitem[Croton et al.(2004)]{2004mnras.352.1232C} Croton, D.~J., et al.\ 
2004, \mnras, 352, 1232 
\bibitem{} Doroshkevich, A. G., 1970, Astrofizica, 6, 581 [Astrophysics 6,
  320]
\bibitem{} Doroshkevich, A. G., Tucker, D. L., Lin, H., Turchaninov, V., Fong,
  R., 2001, MNRAS 322, 369
\bibitem{} Gott, J. R. III, Melott, A. L., Dickinson, M., 1986, ApJ 306, 341
\bibitem[Gott et al. 2005]{Gott05} Gott, J.R.I., Juri{\'c},
M., Schlegel, D., Hoyle, F., Vogeley, M., Tegmark, M., Bahcall, N.,
Brinkmann, J., 2005, \apj, 624, 463
\bibitem{}  Hanami, H, 2001, MNRAS, v 327, pp 721-738
\bibitem[Hikage et al.(2006)]{2006apj...653...11H} Hikage, C., Komatsu, E., 
\& Matsubara, T.\ 2006, \apj, 653, 11
\bibitem[Hoffman \& Shaham(1982)]{1982ApJ...262L..23H} Hoffman, Y., \& 
Shaham, J.\ 1982, \apjl, 262, L23
\bibitem{} Icke, V., 1984, MNRAS, 206, 1
\bibitem{} Jost, J., 2002, Riemannian Geometry and Geometric Analysis
  (Springer, third edition)
\bibitem{} Kerscher, M., 2000, Lecture Notes in Physics 554, 36
\bibitem[Kulkarni et al.(2007)]{2007MNRAS.378.1196K} Kulkarni, G.~V., 
Nichol, R.~C., Sheth, R.~K., Seo, H.-J., Eisenstein, D.~J., \& Gray, A.\ 
2007, \mnras, 378, 1196
 \bibitem[Lorensen, W. and Harvey E. (1987)]{marchingcube} Lorensen, William and Harvey E. Cline. Marching Cubes: A High Resolution 3D Surface Construction Algorithm. Computer Graphics (SIGGRAPH 87 Proceedings) 21(4) July 1987, p. 163-170)
 \bibitem{} Milnor, J., 1963, Morse Theory (Princeton University, Princeton,
  NJ)
\bibitem[Novikov et al.(2006)]{2006MNRAS.366.1201N} Novikov, D., Colombi, 
S., \& Dor{\'e}, O.\ 2006, \mnras, 366, 1201
\bibitem{} Peacock John. A. 1998. Cosmological Physics Cambridge Astrophysics.  
\bibitem{} Peebles, P. J. E., 1980, The Large-Scale Structure of the Universe
  (Princeton Univ. Press, 1980)
  \bibitem[Pichon et al. 2007b]{letterAPtest}      Pichon, C., \,Sousbie, T.  Prunet, S., Courtois, H. , Colombi, S.,  Devrient, J.  {\it in preparation}
\bibitem[Pichon \& Aubert(2006)]{2006MNRAS.368.1657P} Pichon, C., \& 
Aubert, D.\ 2006, \mnras, 368, 1657 
\bibitem[Platen et al.(2007)]{2007arXiv0706.2788P} Platen, E., van de 
Weygaert, R., \& Jones, B.~J.~T.\ 2007, ArXiv e-prints, 706, 
arXiv:0706.2788 
\bibitem{} Sahni, V., Sathyaprakash, B. S., Shandarin, S. F., 1998, ApJ 495,
\bibitem[Sheth \& Sahni(2005)]{2005astro.ph..2105S} Sheth, J.~V., \& Sahni, 
V.\ 2005, ArXiv Astrophysics e-prints, arXiv:astro-ph/0502105
\bibitem[Springel(2005)]{2005MNRAS.364.1105S} Springel, V.\ 2005, \mnras, 
364, 1105
\bibitem[Szapudi et al.(2005)]{2005ApJ...631L...1S} Szapudi, I., Pan, J., 
Prunet, S., \& Budav{\'a}ri, T.\ 2005, \apjl, 631, L1 
\bibitem[Sousbie, PhD Thesis, 2006]{TheseSousbie} Sousbie, T., PhD thesis, 2006.  http://hal-insu.archives-ouvertes.fr/
\bibitem[Sousbie et al. 2007a]{SouSkLet} Sousbie, T., Pichon, C.,  Courtois, H., Colombi, S. Novikov, D. submitted to ApJLett. (arXiv:astro-ph/0602628)
%

%


\end{thebibliography}
\end{document}